\documentclass[a4paper]{article}
\usepackage{JCAP}

\usepackage[T1]{fontenc}
\usepackage{lmodern}

\usepackage{graphicx}	
\usepackage{amsmath}	
\usepackage{amssymb}	
\usepackage{comment}
\usepackage{caption}
\usepackage{subcaption}

\usepackage{pdflscape}
\usepackage{hyperref}
\usepackage{microtype}
\usepackage{color}
\usepackage[utf8]{inputenc}

\defcitealias{Khaire23}{Paper I}
\defcitealias{Gjergo2025}{GK25}

\graphicspath{{./}{figures/}}

\title{Non-Primordial Contribution to the Cosmic Microwave Background}

\author[1]{Akshith Asundi\thanks{Corresponding author: \href{mailto:akshith.asundi@gmail.com}{\texttt{akshith.asundi@gmail.com}}}}
\author[1]{Vikram Khaire}
\affil[1]{Indian Institute of Technology, Tirupati, 517619, India}

\date{\today}

\begin{document}
\maketitle

\begin{abstract}
The cosmic microwave background (CMB) is interpreted as relic thermal radiation from the epoch of recombination at $z \approx 1090$. Recent work by Gjergo \& Kroupa \citep{Gjergo2025} (hereafter \citetalias{Gjergo2025}) has challenged the assumption that this radiation is entirely primordial, proposing that dust-enshrouded starbursts in the present-day progenitors of massive early-type galaxies at $z \sim 15$-$20$ contribute to the CMB at the percent level of the present-day CMB energy density. Existing tests of this scenario rely on the CMB monopole spectrum; here we present the first test against the CMB anisotropy power spectrum, which probes the dynamical imprint of the radiation content on the sound horizon and acoustic peaks independently of the monopole's absolute calibration. We introduce a single dimensionless emissivity parameter $\varepsilon_{\rm CMB}$ that rescales the primordial photon energy density above a transition redshift $(z_t=17)$ treating the non-primordial component as isotropic and Planckian, and implement it self-consistently. Constraining the model  with the Planck 2018 TT, TE, EE, low-$\ell$ polarisation and lensing likelihoods combined with baryon acoustic oscillation measurements, we find a percent-level dust contribution to be fully consistent with the data. When we restrict $\varepsilon_{\rm CMB} \leq1$, the data place a 68 (95) per cent lower bound $\varepsilon_{\rm CMB} \gtrsim 0.971\ (0.953)$, permitting a non-primordial contribution of up to $\sim 3$ ($\sim 5$) per cent of the CMB energy density. When sampled with $\varepsilon_{\rm CMB}$ as a free parameter, the data yield a well-resolved posterior with $\varepsilon_{\rm CMB} = 1.0104^{+0.025}_{-0.024}$. The $1.4$ per cent contribution conservatively estimated by \citetalias{Gjergo2025} lies comfortably within the 1$\sigma$ region of both cases and is fully consistent with current observations, whereas a contribution approaching the full CMB energy density ($\varepsilon_{\rm CMB} \to 0 $) is excluded at high significance ($42\sigma$). Treating $z_t$ as a free parameter returns a posterior that is flat across $5 \lesssim z_t \lesssim 50$, which indicates the anisotropy spectrum constrains how large a dust contribution can be, but is insensitive to when it occurred. The standard $\Lambda$CDM parameters  including $H_0$ and $\sigma_8$ are recovered without significant shifts, showing that current CMB precision cosmology has sufficient room to accommodate a dust contribution at the level predicted by \citetalias{Gjergo2025} without detectable consequence for parameter inference. Our $\varepsilon_{\rm CMB} - z_t$ framework developed here is not tied to this scenario and provides a general, reusable test for any model in which part of the CMB is non-primordial.
\end{abstract}

\keywords{cosmic microwave background, grey-body radiation, CMB foregrounds, cosmological parameters, Planck 2018}

\section{Introduction} \label{sec:intro}

The cosmic microwave background (CMB) is the most precisely characterised
observable in cosmology and provides the empirical foundation of the standard
cosmological model.
Since its serendipitous discovery as an isotropic excess antenna temperature
\citep{PenzasWilson1965}, the CMB has been established as a near-perfect
blackbody: the COBE (Cosmic Background Explorer)/FIRAS (Far-Infrared Absolute Spectrophotometer) instrument measured its monopole spectrum to be
thermal to better than one part in $10^{4} - 10^{5}$ across the frequency
range $60$-$630\,$GHz \citep{Mather1994, Fixsen1996}, with a present-day
temperature $T_{0} = 2.7255 \pm 0.0006\,$K \citep{Fixsen2009} and its redshift evolution \citep{Srianandmain, Noterdaeme-Srianand}.
The small-amplitude anisotropies in its temperature and polarisation fields,
mapped successively by COBE, WMAP(Wilkinson Microwave Anisotropy Probe) 
\citep{Bennett2013}, and \textit{Planck} \citep{PlanckCollaboration2020} (referred to as \textit{Planck} 2018),
encode the initial conditions and subsequent evolution of the Universe
\citep[see for e.g.,][]{Sunyaev1970, Peebles1968}.
Interpreted within the six-parameter spatially-flat $\Lambda$CDM framework,
the \textit{Planck} 2018 measurements of the angular power spectra constrain the cosmological parameters at sub-percent precision \citep{PlanckCollaboration2020}, an achievement that underpins the current era of precision cosmology. 

In the standard picture, the entire CMB monopole is primordial: relic thermal radiation released at the epoch of photon decoupling at $z \approx 1090$,
approximately $380{,}000$ years after the Big Bang, when the primordial plasma
cooled sufficiently for electrons and protons to recombine into neutral hydrogen
and the Universe became transparent to radiation.
The microwave sky observed today, however, is not strictly primordial.
It is contaminated by Galactic emission such as thermal dust, synchrotron, and
free-free radiation, and by extragalactic backgrounds, the modelling and removal
of which are a central task of any CMB analysis \citep{Planck2018components}.
Of particular relevance here is the cosmic infrared background (CIB), the
integrated and redshifted thermal emission of dusty star-forming galaxies
accumulated over cosmic history. The CIB was first established observationally
as a diffuse far-infrared/submillimetre excess in the COBE data
\citep{Puget1996}, with its spectrum and angular properties subsequently
characterised in detail through further COBE analyses \citep{Fixsen1998,
HauserDwek2001}; more recent work has modelled its build-up through
galaxy-population synthesis and deep extragalactic surveys
\citep{ Bethermin2012, CaseyNarayananCooray2014}.
The CIB demonstrates that the dust reprocessing of stellar light builds up a
diffuse extragalactic background whose bolometric energy density is comparable
to that of the integrated optical and ultraviolet starlight 
\citep[see for e.g][]{Khaire15ebl, KS19}.
An analogous, but far earlier, episode of dust-reprocessed star formation could in
principle contribute not just to the far-infrared background but, once redshifted
by the expansion of the Universe, to the microwave band occupied by the CMB itself.

This possibility has recently been investigated quantitatively
by E. Gjergo \& P. Kroupa 2025 \citep{Gjergo2025} (hereafter GK25), who propose that a non-negligible 
fraction of the observed microwave background may be of astrophysical rather than 
primordial origin. In their scenario, the progenitors of present-day massive early-type galaxies 
(ETGs) undergo intense, dust-enshrouded starbursts at $z \sim 15$-$20$, 
consistent with the monolithic, early-collapse picture of ETG formation \citep{Eggen1962,Larson1974}. 
The bolometric luminosity of these starbursts was absorbed and thermalised by the 
surrounding dust at temperatures $T_{\rm dust} \sim 50\,$K and re-emitted as a 
modified blackbody (grey-body) spectrum which, redshifted by the subsequent expansion, 
peaks in the microwave today. \citetalias{Gjergo2025} estimate that this 
mechanism contributes at least $1.4$ per cent of the present-day CMB energy density, 
with the upper bound potentially much larger depending on assumptions about the 
completeness of the ETG progenitor population.

The scenario has acquired particular interest in the current era of the
\textit{James Webb} Space Telescope (JWST), which has revealed a population of 
surprisingly massive and morphologically evolved galaxies at high redshift. 
The initial detections \citep{Labbe2023} were followed by further characterisation of the
population \citep{Xiao2024}, some of which are difficult to reconcile with the assembly 
histories expected in $\Lambda$CDM \citep{BoylanKolchin2023}. 
This is consistent with independent evidence that galaxy demographics 
more broadly strain the standard picture \citep{Haslbauer2022}. 
A recent age-dating analysis also finds individual JWST-selected galaxies in tension with the $\Lambda$CDM age of the Universe at their redshift \citep{LopezCorredoira2026}. 
Proposed resolutions include early dark energy \citep{Shen2024EarlyDarkEnergy} 
and a reassessment of the systematic uncertainties in the stellar mass  estimates 
themselves \citep{KrishnanAbazajian2026}. Other responses to the same JWST-driven 
tension have pursued entirely
different mechanisms rather than a non-primordial CMB. For example, \citep{Gupta2023CCCTL}
proposes a hybrid tired-light and covarying-coupling-constants cosmology
motivated by the same population of unexpectedly massive early galaxies,
stretching the age of the Universe to accommodate their formation; notably,
that work explicitly flags CMB isotropy as a difficulty for tired-light models
without providing a quantitative test. A modified-gravity alternative along
similar lines has been explored by \citep{Moffat2024MOG}, while
\citep{vanPutten2023FastFurious} approaches the same tension from the
standpoint of early structure growth rates. Together, these proposals show
that the JWST early-galaxy population has motivated several structurally
distinct departures from $\Lambda$CDM, of which the \citetalias{Gjergo2025}
grey-body scenario is one. Our analysis directly addresses that one scenario,
though the framework we develop is not restricted to it (see
Section~\ref{sec:discussion}).

The \citetalias{Gjergo2025} scenario has already been confronted 
with the CMB monopole flux measurements in a recent research note
\cite{Corredoira2026_resnote}. Fitting the COBE-FIRAS spectrum with a
superposition of a primordial blackbody and redshifted dust emission, that
analysis finds that for realistic dust properties 
(spectral index $\beta \geq 0.5$ and a finite spread $\sigma_d$ in the 
epoch of ETG progenitor formation) the dust contamination is limited to 
$<1.3\%$ at 95 per cent confidence, marginally below the conservative 
$1.4\%$ estimate of \citetalias{Gjergo2025}. This bound, however, rests on 
the \emph{chromaticity} of the dust, i.e., 
in the spectrally degenerate limit of Planckian dust emission
(with $\beta=0$) released at a single epoch, the redshifted dust and 
primordial radiation become indistinguishable and the monopole permits a 
contribution as large as the full CMB energy density. The monopole spectrum 
alone thus cannot exclude a fully non-primordial CMB.

The CMB anisotropy power spectrum provides a complementary handle that is
insensitive to the spectral shape of the dust. Instead, it responds to the 
\emph{dynamical} imprint of the photon energy density on the acoustic peak 
structure, an achromatic effect that persists however closely the dust 
mimics a blackbody. Testing the \citetalias{Gjergo2025} scenario against 
the anisotropy spectrum is therefore not merely a further consistency check 
but a qualitatively independent probe.
Two questions follow from such a test: first, is a dust contribution at the percent 
level or below compatible with the highly precise measurements of the acoustic peaks 
in the \textit{Planck} angular power spectra?
Second, if such a component is present but not modelled, does it bias the
inference of the standard cosmological parameters?
The latter question has far-reaching consequences, as also noted by
\citetalias{Gjergo2025}.
A modification to the photon energy density alters the sound horizon
$r_{s} \propto \rho_{\gamma}^{-1/2}$, the photon-baryon ratio that governs
the relative heights of the odd and even acoustic peaks, and the redshift of
matter-radiation equality, and it therefore propagates into derived quantities
such as the Hubble constant $H_{0}$ and the matter clustering amplitude
$\sigma_{8}$.
Given the persistent tensions between early- and late-Universe determinations
of these quantities, such as the Hubble tension \cite{Riess2022,Wong2020H0LiCOW,Verde2019,DiValentino2021}
and the $\sigma_{8}$ tension \cite{Heymans2021KiDS1000,DESCollaboration2022Y3,DiValentino2021S8}, 
any foreground that has not been modelled but is capable of mimicking a 
shift in the radiation content warrants careful quantification.
Moreover, whereas \citetalias{Gjergo2025} frame their estimate as a budget on 
the present-day CMB energy density, the anisotropy spectrum offers a distinct 
observational handle: it probes the dynamical effect of the radiation content on 
the sound horizon and peak structure, rather than requiring an independent 
accounting of the monopole's absolute calibration.

In this paper, we address these questions directly, performing the first
observational test of the \citetalias{Gjergo2025} scenario against
the CMB anisotropy power spectrum. In this scenario, the dust emission
contributing to CMB energy density requires that the CMB radiation is not
entirely primordial, and therefore, we parametrise the non-primordial
contribution through a single dimensionless emissivity parameter,
$\varepsilon_{\rm CMB}$. This $\varepsilon_{\rm CMB}$ parameter rescales the
primordial photon energy density ($\rho_{\gamma}$) above a transition redshift
$z_{t}$, motivated by the \citetalias{Gjergo2025} peak epoch of ETG starburst
formation. Because the dust emission originates from high-redshift
($z_{t} \sim 15$-$20$) sources distributed across the observable Universe, we
treat it, to first approximation, as isotropic on \textit{Planck} angular
scales -- an assumption we revisit and qualify in
Section~\ref{sec:limitations}, where we note that \citetalias{Gjergo2025}
themselves point out a possible hemispherical asymmetry in the ETG progenitor
population. Under this simplifying assumption, we model its principal effect
as a rescaling of the homogeneous photon energy density rather than as an
additional source of anisotropy.

The $\varepsilon_{\rm CMB}$ modification is implemented self-consistently 
at the Fortran source level in
the Boltzmann code \textsc{camb} \citep{Lewis2000, Howlett2012}, so that
$\varepsilon_{\rm CMB}$ and $z_{t}$ are native parameters of the cosmological
model, propagated through the background expansion, the perturbation evolution,
the sound horizon, and the derived epochs of recombination and baryon drag.
We constrain the model through a Markov Chain Monte Carlo analysis using
\textsc{cobaya} \citep{Torrado2021}, combining the \textit{Planck} 2018
high-$\ell$ and low-$\ell$ TT, TE, and EE likelihoods
\citep{PlanckLikelihood2020}, the \textit{Planck} CMB lensing likelihood
\citep{PlanckLensing2020}, and baryon acoustic oscillation (BAO) measurements from
the 6dF Galaxy Survey \citep{Beutler2011}, the SDSS DR7 {Sloan Digital Sky Survey Data Release 7} Main Galaxy Sample
\citep{Ross2015}, and the SDSS-III BOSS {Baryon Oscillation Spectroscopic Survey} DR12 consensus analysis \citep{Alam2017}.

The main goal of the paper is threefold: to determine how much non-primordial 
contribution to the CMB (quantified by $1- \varepsilon_{\rm CMB}$) 
can be accommodated by the current data while remaining
consistent with $\Lambda$CDM; to quantify any resulting shifts in the inferred
cosmological parameters, and to investigate whether the constraint depends on
the assumed epoch of the transition $z_{t}$. 
We measure  $\varepsilon_{\rm CMB} = {1.0104^{+0.025}_{-0.024}}$ and 
\citetalias{Gjergo2025}'s conservative estimate of dust contribution of 1.4\% 
can be easily accommodated, whereas the most drastic case of dust contribution 
approaching full CMB energy density is ruled out by 42$\sigma$. 
The cosmological parameters do not shift from the $\Lambda$CDM baseline 
(where  $\varepsilon_{\rm CMB} =1$) significantly, and the results are 
insensitive to the epoch of dust emission from $5<z_{t}<50$. 

The paper is organised as follows.
Section~\ref{sec:model} describes the grey-body emissivity model and its
physical motivation.
Section~\ref{sec:method} details the numerical implementation in \textsc{camb}
and the MCMC methodology used to constrain the model.
Section~\ref{sec:results} presents the parameter constraints and compares the
grey-body extension with standard $\Lambda$CDM.
Section~\ref{sec:discussion} discusses the implications of our results, and
Section~\ref{sec:conclusions} summarises our conclusions.

\section{Model}
\label{sec:model}

\subsection{Physical motivation}
\label{sec:motivation}

The standard interpretation of the CMB treats the observed microwave background as a 
pure primordial blackbody, originating entirely from photon decoupling at $z \approx 
1090$. \citetalias{Gjergo2025} have challenged this assumption by proposing that a 
non-negligible fraction of the observed CMB energy density arises from the dust 
emission. In this picture, the progenitors of present-day ETGs underwent intense, 
dust-enshrouded starbursts at redshifts $z_{t} \sim 15$-$20$. 
The bolometric luminosity of these starbursts was fully absorbed and thermalised by 
surrounding dust at temperatures of order $T_{\rm dust} \sim 50\,\rm K$, and 
subsequently redshifted into the microwave regime by the expansion of the Universe. 
The resulting contribution is a foreground component superimposed on the primordial 
CMB at the redshift $z_{t}$.  \citetalias{Gjergo2025} estimate that this mechanism can account for at least 1.4\% of the present-day CMB energy density, with the upper bound potentially approaching 100\% depending on model assumptions regarding the completeness of the ETG progenitor population.


Whether a dust contribution at even the sub-percent level is compatible 
with the CMB anisotropy power spectrum, and what effect such a component 
would have on the inferred cosmological parameters if left unmodelled, 
has not previously been addressed; this is the central question of the 
present paper. Since the grey-body component arises from high-redshift 
dust emission that is largely isotropic on large scales, its primary 
effect is to modify the photon energy density rather than to introduce 
anisotropic fluctuations, and we model its impact accordingly as an 
effective rescaling of the homogeneous photon energy density as 
described in the next Section~\ref{sec:param}.

\subsection{Emissivity parametrisation}
\label{sec:param}

\begin{figure}
    \centering
    \includegraphics[width=0.95\columnwidth]{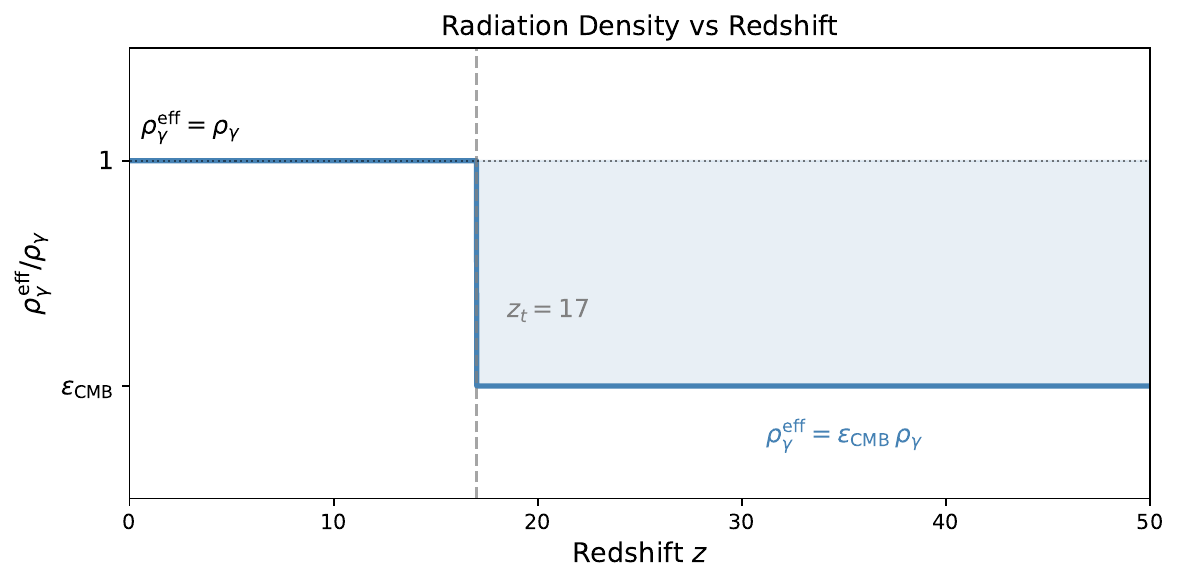}
    \caption{Schematic of the $\varepsilon_{\rm CMB}$ model (equation~\ref{eq:rho_eff}). The curve shows the ratio of the effective to the standard photon energy density, $\rho_\gamma^{\rm eff}/\rho_\gamma$, as a function of redshift for the fiducial transition redshift $z_t = 17$ (vertical dashed line) and an illustrative deficit $\varepsilon_{\rm CMB} < 1$ (shaded region). Above $z_t$, the ratio is suppressed by the constant factor $\varepsilon_{\rm CMB}$, representing an early epoch in which some of the photon energy budget is supplied by redshifted dust emission rather than the primordial plasma; below $z_t$, the ratio returns to unity by construction, so the present-day photon bath is treated as the fixed blackbody endpoint of this evolution. The horizontal dotted line at $\rho_\gamma^{\rm eff}/\rho_\gamma = 1$ marks the standard $\Lambda$CDM case, $\varepsilon_{\rm CMB} = 1$, for reference. It is this step function, propagated through \textsc{camb}, that is fit to the Planck 2018 and BAO data in Section~\ref{sec:results}.}
    \label{fig:step_function}
\end{figure}

We parametrise the dust contribution through a single dimensionless emissivity parameter, $\varepsilon_{\rm CMB}$, which rescales the standard photon energy density above a transition redshift $z_t$. The effective photon energy density entering the cosmological evolution equations is taken to be
\begin{equation}
    \rho_\gamma^{\rm eff}(z) = 
    \begin{cases} 
        \varepsilon_{\rm CMB}\, \rho_\gamma(z) & z > z_t , \\ 
        \rho_\gamma(z) & z \leq z_t \,,
    \end{cases}
    \label{eq:rho_eff}
\end{equation}
where
\begin{equation}
    \rho_\gamma(z) = \frac{\pi^2}{15}\, {T_{\rm CMB}(z)}^4
    \label{eq:rho_blackbody}
\end{equation}
\begin{equation}
    {T_{\rm CMB}(z)}= T_{\rm 0}(1+z)
\end{equation}
is the standard blackbody photon energy density at temperature $T_{\rm 0} = 2.7255\,\rm K$ \cite{Fixsen2009}. The step-function form of equation~(\ref{eq:rho_eff}) is illustrated in Fig.~\ref{fig:step_function} with the $z_t = 17$ for $\varepsilon_{\rm CMB} \leq 1$.

The case $\varepsilon_{\rm CMB} = 1$ recovers standard $\Lambda$CDM exactly (as tested in Section \ref{sec:baseline}).
Values $\varepsilon_{\rm CMB} < 1$ correspond to a deficit of primordial 
photons relative to the full observed blackbody 
(tested in Section \ref{sec:run1}, our Run 1),
as in the grey-body CMB  scenario implied by \citetalias{Gjergo2025}, 
i.e., a fraction of the present-day CMB energy density 
originates from astrophysical dust emission rather 
than the primordial plasma, so the purely primordial contribution is 
grey-body at $z > z_t$. 
Values $\varepsilon_{\rm CMB} > 1$ would instead correspond to an excess 
of primordial photons above the observed blackbody, a 
\textit{super-blackbody} CMB,  which has no direct physical motivation in the present 
context (tested in Section 4.3, our Run 2). 
In both cases, the modification to the cosmological evolution equations 
is confined by construction to epochs above $z_t$; the present-day 
photon bath ($z<z_t$) is treated as the fixed blackbody endpoint of this 
evolution and is not itself further modified by the equations. This is a 
statement about the dynamics rather than about the physical origin of 
today's photons: in the \citetalias{Gjergo2025} picture the observed CMB 
at $z=0$ is itself a superposition of primordial and dust emission, so 
the late Universe ($z_t < 1090$) is precisely where the non-primordial 
contribution resides.

In this paper, the transition redshift $z_t$ is motivated by the epoch 
of peak star formation in the ETG progenitor scenario. 
\citetalias{Gjergo2025} place the luminosity peak of the starburst 
population at $z \approx 15$-$20$, corresponding to approximately 180-
270 Myr after the Big Bang.
We therefore adopt $z_t = 17$ as the 
fiducial value in our baseline analysis, broadly centred within this 
range. The sensitivity of our results to this choice is explored in a 
companion run in which $z_t$ is treated as a free parameter 
(in Section~\ref{sec:zdep}, our Run 3).

The sharp step-function transition at $z_t$ is a minimal 
phenomenological approximation. It is sufficient for the present purpose 
where we are testing the impact of this non-blackbody radiation component on CMB 
parameter inference without committing to a detailed model for its 
astrophysical production history. In the baseline analysis, $z_t$ is 
held fixed, so that $\varepsilon_{\rm CMB}$ constitutes the sole 
additional degree of freedom beyond the standard six $\Lambda$CDM 
parameters.

We emphasise that equation~(\ref{eq:rho_eff}) modifies only the
\emph{energy density} of the radiation field, not its spectral shape, i.e., 
the photon distribution entering the Boltzmann hierarchy, Thomson
scattering, and the recombination calculation is treated as Planckian
throughout, with $\varepsilon_{\rm CMB}$ rescaling its integrated
energy density above $z_t$. The parametrisation is therefore
achromatic by construction. This is a controlled approximation rather
than an arbitrary one: the redshifted dust temperature in the \citetalias{Gjergo2025}
scenario nearly coincides with the observed CMB temperature
($\Delta T/T_0 \approx 2$ per cent; Section~5.4.1), 
so spectral corrections to the background and perturbation dynamics
enter only at second order in $\Delta T/T_0$. Constraints on the
spectral \emph{shape} of any non-primordial component are the domain
of the CMB monopole (Section~\ref{sec:discussion}), and are fully
complementary to the energy-density constraint derived here.

Note that our approach of modifying the CMB energy density above a transition
redshift $z_t$ by rescaling it with the parameter $\varepsilon_{\rm CMB}$ offers
a more general way of quantifying the degree to which the CMB spectrum resembles
a grey-body ($\varepsilon_{\rm CMB} < 1$, a deficit) or a super-blackbody
($\varepsilon_{\rm CMB} > 1$, an excess) at the epoch at which it was released,
after recombination (or even before it, depending on the value of $z_t$).
Should the data definitively favour such a departure, a physical model capable
of accounting for it would be required. While this is presently motivated by the
model of \citetalias{Gjergo2025}, the parametrisation is not exclusive to it and
carries implications for other non-standard cosmological scenarios proposed in
the literature \cite{Zwicky1929TiredLight,MeliaShevchuk2012RhCT,HoyleBurbidgeNarlikar1993QSSC}.

\section{Method}
\label{sec:method}

\subsection{CAMB implementation}
\label{sec:camb_impl}

The $\varepsilon_{\rm CMB}$ model described in Section~\ref{sec:model} is 
implemented by modifying the Boltzmann code \textsc{camb} \cite{Lewis2000, 
Howlett2012}. The modification is introduced at the level of the \textsc{Fortran} 
source code, so that $\varepsilon_{\rm CMB}$ and $z_t$ are native parameters of the 
cosmological model rather than post-processing corrections. The six standard 
cosmology parameters and other notations are explained further in Table~\ref{tab:params_description}.

The effective photon energy density (equation~\ref{eq:rho_eff}) is propagated 
consistently through all relevant \textsc{camb} routines. Specifically, the 
following locations are modified:

\begin{enumerate}
    \item \textit{Background expansion rate} (\texttt{dtauda}$^{\star}$\footnote{The superscript $\star$ denotes {\sc Fortran} routines in \textsc{camb}}): The conformal time integrand $\mathrm{d}\tau/\mathrm{d}a = 1/(a^2 H)$, where $a$ is the scale factor and $H$ is the Hubble parameter, is computed from the total energy density. Since $\rho_\gamma$ contributes directly to the Hubble parameter, rescaling it by $\varepsilon_{\rm CMB}$ modifies the expansion history at all epochs above $z_t$, affecting the comoving distance to recombination and hence all angular scales in the power spectrum.
    \item \textit{Perturbation evolution} (\texttt{derivs}$^\star$, three instances within the routine): 
    The photon energy density entering the Boltzmann hierarchy for scalar, tensor, and vector perturbations is rescaled above $z_t$ in each of the three corresponding blocks of \textsc{camb}'s \texttt{derivs} subroutine. This ensures that the radiation driving of metric perturbations and the photon-baryon plasma oscillations remain consistent with the rescaled expansion history computed in \texttt{dtauda} above.

    \item \textit{Sound horizon} (\texttt{dsound\_da\_exact}$^\star$, 
    \texttt{dsound\_da\_approx}$^\star$): The photon-baryon ratio $(r=3\rho_b / 4\rho_\gamma)$, which sets the sound speed 
    $c_s = c/\sqrt{3(1+r)}$ and hence the sound horizon $r_s = \int c_s\, \mathrm{d}\tau$, is computed using $\rho_\gamma^{\rm eff}$. The \texttt{dsound\_da\_approx}$^\star$ function is used internally by our Markov Chain Monte Carlo (MCMC) sampler \textsc{cobaya} 
    to convert the sampled parameter $\theta_{\rm MC}$ to $H_0$ (see Table~ \ref{tab:params_description});
    updating it ensures the derived Hubble constant is consistent with the modified radiation content.

    \item \textit{Matter-radiation equality and decoupling} (\texttt{get\_zstar}$^\star$, \texttt{get\_zdrag}$^\star$): The redshift of matter-radiation equality $z_{\rm eq} \propto \rho_m/\rho_\gamma$ is updated directly with $\rho_\gamma^{\rm eff}$, since it governs the relative heights of the acoustic peaks through the early integrated Sachs-Wolfe effect \cite{SachsWolfe1967}.
\end{enumerate}

Note that the redshifts of photon decoupling $z_*$ and baryon drag
$z_{\rm drag}$ are determined by the Thomson scattering optical depth, 
which depends on the
ionisation history as computed by \textsc{recfast} \citep{Seager2000} 
and the background expansion rate.
\textsc{recfast} computes recombination from the photon temperature $T_\gamma(z) = T_{\rm 0}(1+z)$
and the baryon density, neither of which is rescaled by $\varepsilon_{\rm CMB}$ in our implementation; the ionisation history is therefore unchanged by construction (see Section~\ref{sec:discussion} for the physical justification of this choice). Both $z_*$ and $z_{\rm drag}$ nevertheless respond to $\varepsilon_{\rm CMB}$
indirectly through the modified expansion rate in \texttt{dtauda}$^\star$. 



These implementations ensures that  $\varepsilon_{\rm CMB}$ modification is self-
consistently propagated into all CMB observables, including the angular power 
spectrum, the acoustic peak positions, and the derived parameters $\theta_*$, $r_s$, 
and $H_0$. 
The initial conditions for the perturbation equations are set at 
conformal times corresponding to redshifts $z_{\rm IC} \gg z_t$ for all mode 
wavenumbers $k$ probed by the Planck 2018 likelihood, as discussed in the next 
subsection, so they unconditionally apply the $\varepsilon_{\rm CMB}$ modification. 

\subsection{Likelihoods}
\label{sec:likelihoods}

Our $\varepsilon_{\rm CMB}$ modified model is constrained using the following data sets:

\begin{itemize}
    \item The Planck 2018 high-$\ell$ TT, TE, and EE power spectrum likelihood ($30 \leq \ell \leq 2508$) \cite{PlanckLikelihood2020};
    \item The Planck 2018 low-$\ell$ temperature and polarization likelihoods ($2 \leq \ell \leq 29$) \cite{PlanckLikelihood2020};
    \item The Planck 2018 CMB lensing likelihood \cite{PlanckLensing2020};
    \item BAO measurements from the 6dF Galaxy Survey \cite{Beutler2011}, the SDSS DR7 Main Galaxy Sample \cite{Ross2015}, and the 
    SDSS-III BOSS DR12 consensus analysis \cite{Alam2017}.
\end{itemize}

The combination of Planck 2018 CMB anisotropies with lensing and BAO provides 
complementary sensitivity to the expansion history and the growth of structure. In 
particular, the inclusion of lensing and BAO data helps to break degeneracies 
between $\varepsilon_{\rm CMB}$ and the standard cosmological parameters, especially 
$\omega_c h^2$ (defined in Table \ref{tab:params_description}) and $H_0$, which both 
respond to changes in the effective radiation content. The primary sensitivity to 
$\varepsilon_{\rm CMB}$ comes from the TT, TE, and EE power spectra, which constrain 
the acoustic peak structure and hence the photon energy density at recombination.

\subsection{MCMC sampling and convergence}
\label{sec:mcmc}

Parameter inference is performed using the \textsc{cobaya} sampler \cite{Torrado2021} 
with MCMC sampling. Five parallel chains were used
with adaptive proposal covariance learning enabled. The sampled parameter 
vector consists of the six standard flat-$\Lambda$CDM parameters together with 
$\varepsilon_{\rm CMB}$, as summarised in Table~\ref{tab:params_description}.

\begin{table}
    \centering
    \caption{The six standard flat-$\Lambda$CDM parameters sampled in the MCMC analysis, together with the $\varepsilon_{\rm CMB}$ parameter. The Hubble constant $H_0$ is a derived parameter, computed self-consistently from $\theta_{\rm MC}$ 
    and the sound horizon with the emissivity modification active.}
    \label{tab:params_description}
    \begin{tabular}{lp{4cm}p{5cm}}
    \hline
    Parameter & Definition & Physical role \\
    \hline
    $\theta_{\rm MC}$ & Ratio of the sound horizon at photon decoupling to the comoving angular diameter distance to the last-scattering surface. & Approximate angular scale of the acoustic peaks; primary geometric parameter, related to $H_0$ through the sound horizon. \\[4pt] 
    $\omega_b h^2$ & Baryon density parameter $\Omega_b$ multiplied by $h^2$, where $h = H_0/(100\,{\rm km\,s^{-1}\,Mpc^{-1}})$. & Physical baryon density; controls the baryon-to-photon ratio and hence the relative heights of odd and even acoustic peaks. \\[4pt]
    $\omega_c h^2$ & Cold dark matter density parameter $\Omega_c$ multiplied by $h^2$. & Physical cold dark matter density; sets the epoch of matter-radiation equality and the small-scale damping of the power spectrum. \\[4pt]
    $A_s$ & Amplitude of the primordial curvature power spectrum at a pivot scale $k_0 = 0.05\,{\rm Mpc}^{-1}$. & Amplitude of primordial scalar perturbations; sets the overall normalisation of the power spectrum. \\[4pt]
    $n_s$ & Power-law index of the primordial scalar power spectrum, $P(k) \propto k^{n_s-1}$. & Scalar spectral index; parametrises the tilt of the primordial power spectrum, sensitive to the physics of inflation. \\[4pt]
    $\tau$ & Integrated Thomson scattering optical depth of CMB photons due to reionisation. & Reionisation optical depth; suppresses power at small angular scales, constrained primarily by the low-$\ell$ polarisation spectrum. \\[4pt]
    \hline
    $\varepsilon_{\rm CMB}$ & Dimensionless factor rescaling the photon energy density $\rho_\gamma$ above the transition redshift $z_t$ (equation~\ref{eq:rho_eff}). & The emissivity parameter; rescales the primordial photon energy density above $z_t$. Values $\varepsilon_{\rm CMB} < 1$ correspond to a grey-body CMB; values $\varepsilon_{\rm CMB} > 1$ to a super-blackbody CMB. \\
    \hline
    \end{tabular}
\end{table}

Convergence is assessed using the generalised Gelman-Rubin statistic $R-1$  
\cite{Gelman1992}. All runs are continued until $R-1 < 0.005$, which is more 
stringent than the commonly adopted threshold of $R-1 < 0.01$, to ensure reliable 
posterior estimates in the presence of parameter degeneracies involving 
$\varepsilon_{\rm CMB}$. Where chains from a previous run are available, the 
resulting covariance matrix is used to initialise the proposal distribution for 
subsequent runs, which significantly accelerates convergence. Post-processing of the chains is performed using the \textsc{numpy}
\citep{Harris2020numpy}, \textsc{scipy} \citep{Virtanen2020scipy}, and
\textsc{corner} \citep{ForemanMackey2016corner} Python packages to
obtain marginalised posterior distributions and credible intervals.

\subsection{Prior choices}
\label{sec:priors}

Uniform priors are adopted for all parameters. The prior ranges and reference values 
for the standard $\Lambda$CDM parameters are summarised in 
Table~\ref{tab:priors_lcdm}. For the emissivity parameter we consider two prior 
ranges: a conservative range $\varepsilon_{\rm CMB} \in [0.5, 1.0]$ (we call this Run 1), motivated by 
\citetalias{Gjergo2025}, that allows only a deficit of photons relative to the 
blackbody, and a wider range $\varepsilon_{\rm CMB} \in [0.5, 1.5]$ (we call this Run 2) that additionally 
allows for an excess. Results from both ranges (Run 1 and Run 2) are presented in 
Section~\ref{sec:discussion} to assess the impact of the prior boundary on the 
inferred constraints.

\begin{table*}
    \centering
\caption{Uniform prior ranges and reference values for the standard $\Lambda$CDM  parameters used in the MCMC analysis. The proposal width gives the initial step size  for the adaptive sampler. Prior ranges are chosen to be broad relative to the Planck 2018 constraints \cite{PlanckCollaboration2020}.}
\label{tab:priors_lcdm}
\begin{tabular}{lcccc}
\hline
Parameter & Prior min & Prior max & Reference value & Proposal width \\
\hline
$\theta_{\rm MC}$   & 0.0102              & 0.0107              & 0.01041            & $5\times10^{-6}$ \\
$\omega_b h^2$      & 0.02                & 0.024               & 0.0224             & $1\times10^{-4}$ \\
$\omega_c h^2$      & 0.10                & 0.14                & 0.12               & $1\times10^{-3}$ \\
$A_s$               & $1.5\times10^{-9}$  & $3.5\times10^{-9}$  & $2.1\times10^{-9}$ & $1\times10^{-10}$ \\
$n_s$               & 0.90                & 1.05                & 0.965              & $5\times10^{-3}$ \\
$\tau$              & 0.01                & 0.10                & 0.055              & $5\times10^{-3}$ \\
\hline
$\varepsilon_{\rm CMB}$ (run 1) & 0.5 & 1.0 & 1.0 & 0.01 \\
$\varepsilon_{\rm CMB}$ (run 2) & 0.5 & 1.5 & 1.0 & 0.01 \\
\hline
\end{tabular}
\end{table*}

The standard cosmological parameters are sampled in the native \textsc{camb}/\textsc{cobaya} 
parametrisation. All derived quantities - including $H_0$, $\Omega_m$, $\sigma_8$, 
and the acoustic scale $\theta_*$ - are computed self-consistently from the sampled 
parameters with the emissivity modification active.

%
\begin{figure}
    \centering
    \includegraphics[width=\textwidth]{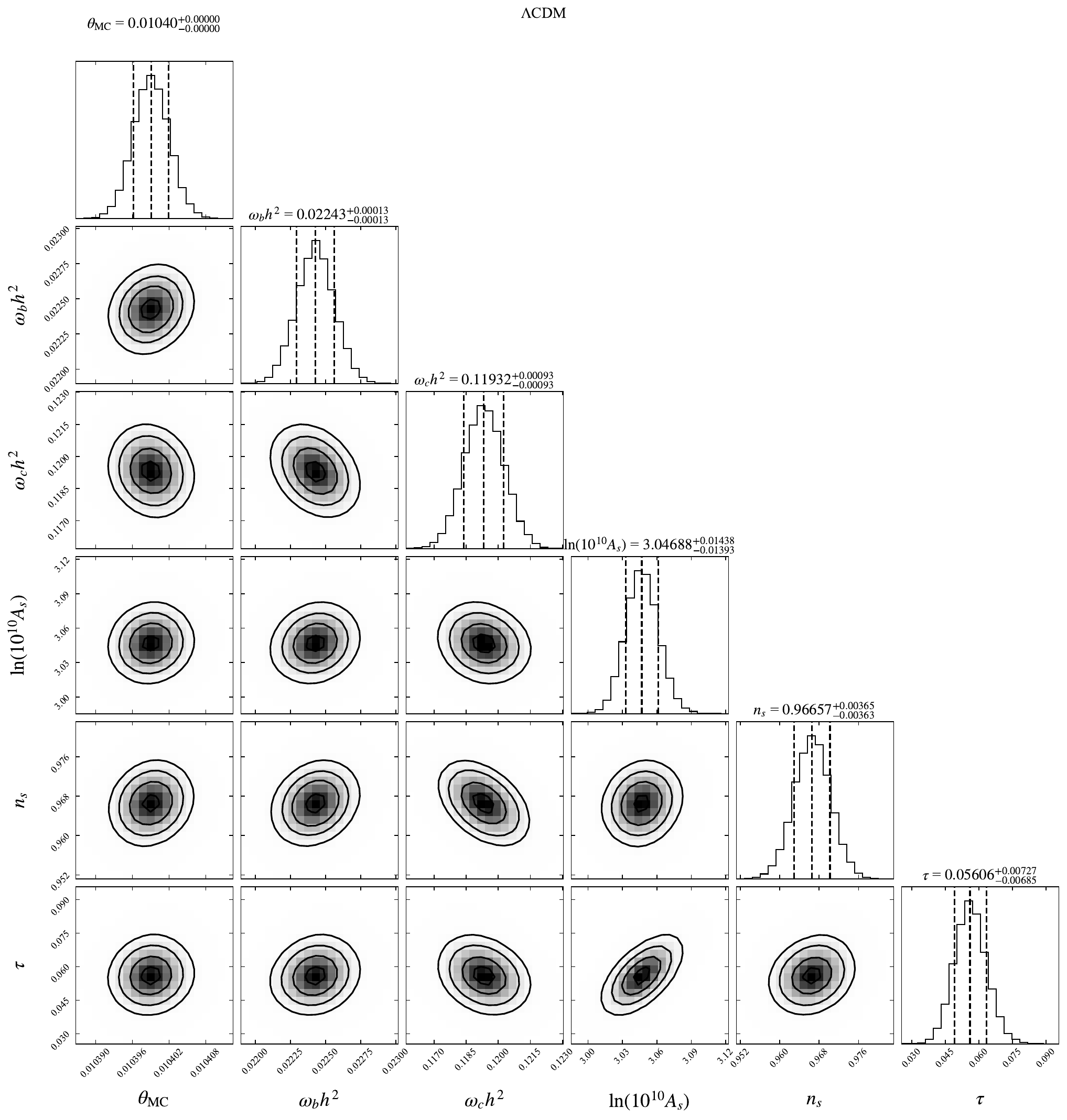}
    \caption{Validation of the modified pipeline against standard $\Lambda$CDM: marginalised posterior distributions for the baseline six-parameter flat $\Lambda$CDM model ($\varepsilon_{\rm CMB}$ fixed to unity), constrained using the Planck 2018 TT, TE, EE, low-$\ell$, lensing, and BAO likelihoods described in Section~\ref{sec:likelihoods}. The diagonal panels show the one-dimensional marginalised posteriors of the six sampled $\Lambda$CDM parameters, $\theta_{\rm MC}$, $\omega_b h^2$, $\omega_c h^2$, $A_s$, $n_s$, $\tau$ (see Table~\ref{tab:params_description}), with dashed vertical lines indicating the median and the bounds of the 68 per cent credible interval. The off-diagonal panels show the corresponding two-dimensional joint posteriors, with contours enclosing the 68 and 95 per cent credible regions. The recovered parameter values agree with the published Planck 2018 results of \citep{PlanckCollaboration2020} (compare Table~\ref{tab:params_summary}), confirming that the modified \textsc{camb} implementation introduces no systematic offset relative to standard $\Lambda$CDM before the $\varepsilon_{\rm CMB}$ correction is switched on in the runs that follow.}
    \label{fig:corner_lcdm}
\end{figure}
\section{Results}
\label{sec:results}

\subsection{Baseline \texorpdfstring{$\Lambda$}{ΛCDM}CDM recovery}
\label{sec:baseline}

Before introducing the $\varepsilon_{\rm CMB}$ parameter explicitly, 
we first validate the pipeline by running the standard six-parameter flat 
$\Lambda$CDM model against the Planck 2018 likelihoods described in 
Section~\ref{sec:likelihoods} (we call this the baseline $\Lambda$CDM run). In this run, $\varepsilon_{\rm CMB}$ is fixed to 
unity, recovering standard $\Lambda$CDM exactly. The marginalised posterior 
distributions are shown in Fig.~\ref{fig:corner_lcdm}. All parameters are recovered in excellent agreement with the Planck 2018
results \citep{PlanckCollaboration2020} (see Table~\ref{tab:params_summary}), including
the derived parameters $H_0$ and $\sigma_8$, confirming that our modification
to the \textsc{camb} implementation and \textsc{cobaya} sampling are
functioning correctly.

The baseline median $\chi^2 = {2794.55^{+6.28}_{-5.14}}$ provides the reference goodness-of-fit against which the extended models are compared.
This baseline run serves a double purpose: it establishes the reference parameter 
values against which shifts in the $\varepsilon_{\rm CMB}$ runs can be measured, and 
it confirms that no systematic offset is introduced by our modified \textsc{camb} 
implementation prior to the activation of the emissivity correction.

\subsection{Grey-body emissivity constraints $\varepsilon_{\rm CMB} < 1$: the physical case}
\label{sec:run1}

We now introduce the emissivity parameter $\varepsilon_{\rm CMB}$ with the prior 
range $[0.5, 1.0]$, corresponding to the physically motivated case in which a 
fraction of the observed CMB energy density has an astrophysical rather than 
primordial origin. We refer to this analysis as Run~1 throughout the remainder of the 
paper.

\begin{figure*}
    \centering
    \includegraphics[width=\textwidth]{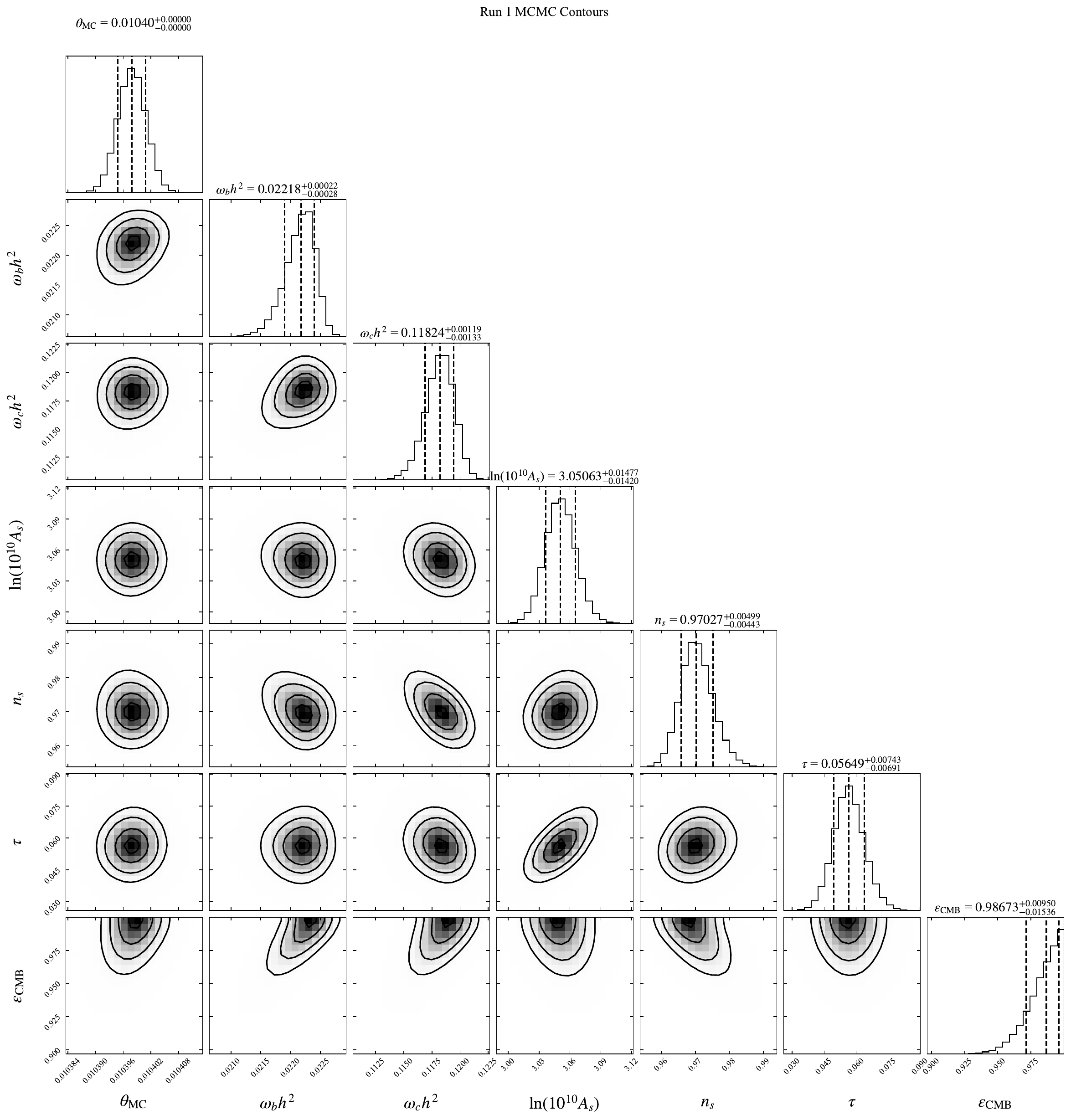}
    \caption{Testing for a physical (deficit-only) dust contribution: marginalised posterior distributions for the $\Lambda$CDM + grey-body CMB model with the one-sided prior $\varepsilon_{\rm CMB} \in [0.5, 1.0]$ (Run~1, fixed transition redshift $z_t = 17$), constrained using the Planck 2018 TT, TE, EE + low-$\ell$ + BAO + lensing likelihoods. This prior range permits only a deficit of primordial photons relative to the observed blackbody, as expected in the \citetalias{Gjergo2025} dust scenario. The diagonal panels show the one-dimensional marginalised posteriors of the six $\Lambda$CDM parameters plus $\varepsilon_{\rm CMB}$, with dashed lines marking the median and the 68 per cent credible interval; note that the $\varepsilon_{\rm CMB}$ posterior (bottom-right panel, also shown enlarged in Fig.~\ref{fig:posterior_run1}) rises monotonically up to the prior edge at unity rather than resolving to an interior peak. The off-diagonal panels show the two-dimensional joint posteriors, with contours enclosing the 68 and 95 per cent credible regions. All six standard cosmological parameters remain consistent with their baseline $\Lambda$CDM values (Fig.~\ref{fig:corner_lcdm}). A strong positive correlation is visible between $\varepsilon_{\rm CMB}$ and both $\omega_b h^2$ and $\omega_c h^2$: this reflects a partial degeneracy in which a lower photon energy density can be partly compensated by lower matter densities, discussed in Section~\ref{sec:run1}.}
    \label{fig:corner_run1}
\end{figure*}

\begin{figure}
    \centering
    \includegraphics[width=0.65\textwidth]{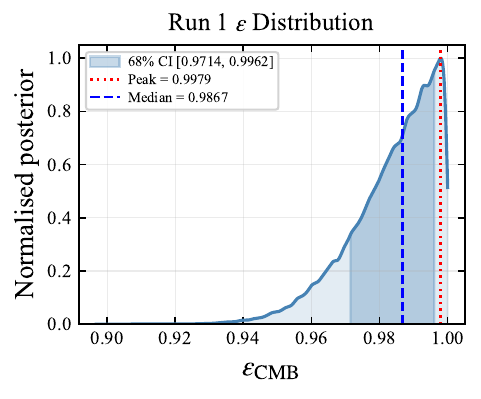}
    \caption{One-dimensional marginalised posterior of $\varepsilon_{\rm CMB}$ for Run~1 (one-sided prior $\varepsilon_{\rm CMB} \in [0.5, 1.0]$, fixed $z_t = 17$; TT, TE, EE + low-$\ell$ + BAO + lensing likelihoods), extracted from the bottom-right panel of Fig.~\ref{fig:corner_run1}. The horizontal axis shows $\varepsilon_{\rm CMB}$ and the vertical axis the normalised posterior density. The posterior rises monotonically all the way to the prior boundary at $\varepsilon_{\rm CMB} = 1$ rather than turning over at an interior value, so this run alone cannot distinguish a genuine preference for $\varepsilon_{\rm CMB} \approx 1$ from an artefact of the imposed upper limit (this ambiguity is resolved in Run~2, Fig.~\ref{fig:posterior_run2}, where the prior is widened to allow $\varepsilon_{\rm CMB} > 1$). The red dotted line marks the posterior peak (at the prior edge) and the blue dashed line marks the posterior median, $\varepsilon_{\rm CMB} = 0.9867$; the shaded region indicates the 68 per cent credible interval, $[0.9713, 0.9963]$, whose lower edge gives the quoted bound on the maximum allowed grey-body deficit. The \citetalias{Gjergo2025} conservative dust-contribution estimate, $\varepsilon_{\rm CMB} \approx 0.986$, falls within this 68 per cent credible interval and close to the posterior median, indicating that this level of contamination is fully consistent with, and not disfavoured by the Planck 2018 data.}
    \label{fig:posterior_run1}
\end{figure}

The marginalised posterior distribution of $\varepsilon_{\rm CMB}$ is shown in 
Fig.~\ref{fig:posterior_run1}, and the full joint posterior including all sampled 
parameters is shown in Fig.~\ref{fig:corner_run1}. The posterior of $\varepsilon_{\rm 
CMB}$ rises monotonically toward the prior boundary at $\varepsilon_{\rm CMB} = 1$, 
suggesting that the data do not favour a departure from standard $\Lambda$CDM
within this prior range. The distribution yields a median of 
$\varepsilon_{\rm CMB} = {0.9867}$, corresponding to a deviation of ${-1.33}$ per 
cent from the blackbody value, with a 68 per cent credible interval of ${[0.9713, 
0.9963]}$, computed from the 16th-84th percentiles of the marginalised posterior 
rather than assumed Gaussian error bars, since the posterior is neither symmetric nor 
Gaussian. \emph{This constitutes a lower limit rather than a detection}, because the 
posterior is not uniform or symmetric about any interior peak: instead it rises 
monotonically and piles up against the prior wall at $\varepsilon_{\rm CMB} = 1$ 
(Fig.~\ref{fig:posterior_run1}). This indicates that the upper end of the credible 
interval is set by the prior boundary rather than by the data converging on a 
preferred value away from it. As a result, the 16th-percentile value is the only edge 
of the interval (away from $\varepsilon_{\rm CMB} = 1$) that carries genuine 
constraining information, and it is this value that we quote as the lower bound 
below: at the 68 per cent credible level, the primordial photon deficit is 
constrained to no more than ${2.9}$ per cent 
($\varepsilon_{\rm CMB} \gtrsim {0.9713}$), and at the 95 per cent credible level to 
no more than approximately ${4.7}$ per cent ($\varepsilon_{\rm CMB} \gtrsim {0.953}$).

The physical origin of this constraint lies in the sensitivity of the CMB acoustic 
peak structure to the photon energy density. 
A modification to $\rho_\gamma$ at $z > z_t$ shifts the sound horizon 
$r_s \propto \rho_\gamma^{-1/2}$, alters the photon-baryon ratio 
$(3\rho_b / 4\rho_\gamma)$ that determines the relative heights of odd and even 
acoustic peaks, and changes the redshift of matter-radiation equality 
$z_{\rm eq} \propto \rho_m/\rho_\gamma$. 
Despite this multi-channel sensitivity, the 2.9 (4.7) per cent bound
arises because a reduction in the primordial photon energy density is partially 
degenerate with the standard cosmological parameters: specifically, a decrease in 
$\varepsilon_{\rm CMB}$ can be partially compensated by a corresponding decrease in 
$\omega_c h^2$, which adjusts $z_{\rm eq}$ and preserves the relative peak heights. 
This degeneracy is visible in the joint posterior of Fig.~\ref{fig:corner_run1} as a 
strong correlation between $\varepsilon_{\rm CMB}$ and $\omega_c h^2$. The inclusion 
of BAO and lensing data limits the extent to which $\omega_c h^2$ can shift to 
compensate, but does not eliminate the degeneracy entirely at the 2.9 (4.7) per 
cent level corresponding to 68 (95) per cent confidence limit.

The resulting shifts (Table~\ref{tab:params_summary}) in the standard cosmological parameters relative to the baseline
are small and statistically insignificant. 
Here we quote each shift in units of $\sigma$, where $\sigma$ denotes the Run~1 
posterior's own uncertainty on that parameter, rather than the baseline uncertainty, 
since it is the Run~1 posterior itself whose consistency with the baseline central 
value is in question. The baryon density $\omega_b h^2$ decreases from ${0.02243}$ to 
${0.02218}$, a shift of $0.00025$; relative to the Run~1 uncertainty of 
$^{+0.00022}_{-0.00028}$, this is $\sim 1.1\sigma$, the only standard-parameter shift 
across any run that exceeds $1\sigma$. The cold dark matter density $\omega_c h^2$ 
decreases from ${0.11932}$ to ${0.11823}$, a shift of ${\sim 0.9\sigma}$, again 
relative to its own Run~1 uncertainty, which is close to, but just under, the $1\sigma$ level.

The estimate of \citetalias{Gjergo2025} is worth returning to here: the dust 
emission from ETG progenitor starbursts is predicted to account for at least 1.4 per 
cent of the present-day CMB energy density, corresponding to $\varepsilon_{\rm CMB} 
\approx 0.986$, and this value lies comfortably within the 68 per cent credible 
interval derived here. It falls close to the posterior median rather than in its 
tail, so a dust-emission foreground at the predicted level is about as 
probable as the standard $\Lambda$CDM value under the current data. 

It is also worth noting the independent consistency check offered by the dust 
temperature: \citetalias{Gjergo2025} derive a thermalisation temperature of $T_{\rm 
dust, ETG, em} \approx 50\,\rm K$ at the formation epoch $z \approx 17$, which when 
redshifted to the present day gives 
$T_{\rm dust, obs} = T_{\rm dust, ETG, em}/(1 + z) \approx 2.8\,\rm K$, strikingly 
close to the measured CMB temperature $T_0 = 2.7255\,\rm K$ \cite{Fixsen2009}. 
This near-coincidence --- if it emerges entirely from galaxy formation physics 
without any CMB priors imposed, rather than being hard-coded into the 
\citetalias{Gjergo2025} model assumptions --- provides additional plausibility to 
the dust component in the CMB scenario independently of the anisotropy constraints. 
This was tested quite rigorously with the CMB monopole spectrum in the research 
note by \cite{Corredoira2026_resnote}
finding that 1.3\% of non-primordial contribution by dust is allowed in the 
physically motivated models of \citetalias{Gjergo2025}.

The goodness of fit is essentially unchanged with respect to the baseline: 
$\chi^2 = 2795.72^{+6.50}_{-5.29}$, a difference of $\Delta\chi^2 = +1.17$ relative 
to standard $\Lambda$CDM, for one additional degree of freedom 
($\varepsilon_{\rm CMB}$). The grey-body extension therefore provides no improvement 
in fit, consistent with the posterior accumulating at the prior wall. 
It simply reflects a genuine limitation of the current data: Planck 2018 has neither the precision to detect a foreground contribution at the $\sim 1$-$2$ per cent level, nor to rule one out.

\subsection{Allowing for Super-blackbody $\varepsilon_{\rm CMB} >1$: the unphysical extension}
\label{sec:run2}

To verify that the behaviour seen in Section~\ref{sec:run1} for 
the $\varepsilon_{\rm CMB}$ posterior (in Fig.~\ref{fig:posterior_run1}) 
is not an artefact of the 
prior boundary at $\varepsilon_{\rm CMB} = 1$, we repeat the analysis with the wider
prior range $\varepsilon_{\rm CMB} \in [0.5, 1.5]$, allowing the posterior to extend 
into the super-blackbody regime $\varepsilon_{\rm CMB} > 1$. We refer to this 
analysis as Run~2 throughout the remainder of the paper. 
The monotonic rise seen in Fig.~\ref{fig:posterior_run1} may reflect the data's 
preference for $\varepsilon_{\rm CMB} \approx 1$; if so, when the prior wall is 
removed, it should resolve into a well-defined Gaussian centred near unity. 

\begin{figure*}
    \centering
    \includegraphics[width=\textwidth]{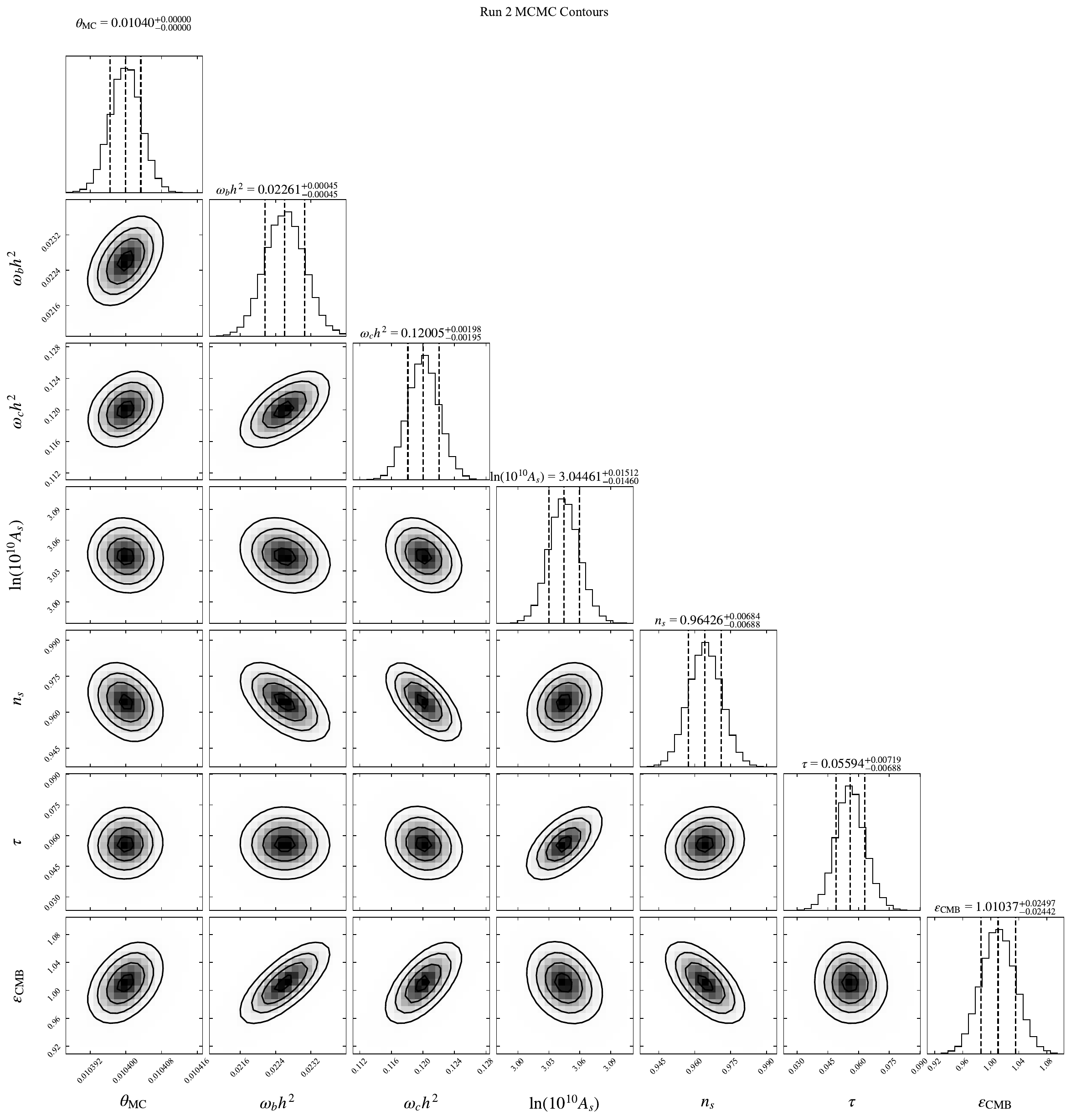}
    \caption{Testing whether the Run~1 result is a prior-boundary artefact: marginalised posterior distributions for the same $\Lambda$CDM + $\varepsilon_{\rm CMB}$ model, but with the wider, two-sided prior $\varepsilon_{\rm CMB} \in [0.5, 1.5]$ (Run~2, fixed $z_t = 17$), which additionally permits a super-blackbody excess ($\varepsilon_{\rm CMB} > 1$). Likelihoods and layout are identical to Fig.~\ref{fig:corner_run1}. With the upper prior wall removed, the $\varepsilon_{\rm CMB}$ posterior (bottom-right panel, enlarged in Fig.~\ref{fig:posterior_run2}) now resolves into a well-defined, symmetric distribution centred close to unity rather than piling up at an edge, confirming that the monotonic rise seen in Run~1 (Fig.~\ref{fig:posterior_run1}) reflects the data's genuine preference for $\varepsilon_{\rm CMB}\approx1$ and is not an artefact of that prior's upper boundary. The six standard cosmological parameters remain consistent with the baseline $\Lambda$CDM values, with somewhat larger uncertainties than in Run~1 owing to the broader $\varepsilon_{\rm CMB}$ prior now being explored on both sides of unity. The same positive correlation between $\varepsilon_{\rm CMB}$ and both $\omega_b h^2$ and $\omega_c h^2$ seen in Run~1 persists here, with Pearson correlation coefficients $r(\varepsilon_{\rm CMB}, \omega_b h^2) = {+0.953}$ and $r(\varepsilon_{\rm CMB}, \omega_c h^2) = {+0.878}$.}
    \label{fig:corner_run2}
\end{figure*}

\begin{figure}[ht!]
    \centering
    \includegraphics[width=0.65\columnwidth]{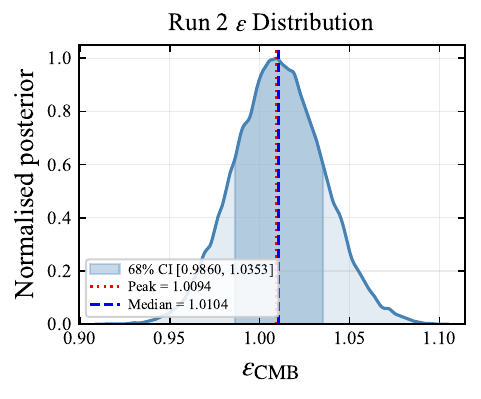}
    \caption{One-dimensional marginalised posterior of $\varepsilon_{\rm CMB}$ for Run~2 (two-sided prior $\varepsilon_{\rm CMB} \in [0.5, 1.5]$, fixed $z_t = 17$; TT, TE, EE + low-$\ell$ + BAO + lensing likelihoods), extracted from the bottom-right panel of Fig.~\ref{fig:corner_run2}, shown for direct comparison with Fig.~\ref{fig:posterior_run1}. Unlike the one-sided Run~1 posterior, which rose monotonically to the prior edge, this distribution is well resolved and closes on both sides, peaking at $\varepsilon_{\rm CMB} = 1.009$ and with median $1.010$; the shaded region marks the 68 per cent credible interval, $[0.986, 1.035]$. The distribution is approximately Gaussian and symmetric about unity, demonstrating that the apparent preference for $\varepsilon_{\rm CMB} \to 1$ seen in Run~1 was a genuine feature of the data rather than a boundary artefact, and that standard $\Lambda$CDM ($\varepsilon_{\rm CMB}=1$) remains fully consistent with the Planck 2018 data at well under $1\sigma$.}
    \label{fig:posterior_run2}
\end{figure}

The complete joint posterior of Run 2 is shown in Fig.~\ref{fig:corner_run2}, with 
the marginalized posterior of $\varepsilon_{\rm CMB}$ shown in 
Fig.~\ref{fig:posterior_run2}. The posterior is indeed a well-resolved Gaussian 
centred at $\varepsilon_{\rm CMB} = {1.0104^{+0.025}_{-0.024}}$, symmetric about 
unity and fully consistent with standard $\Lambda$CDM at well under $1\sigma$. A 
Gaussian fit yields $\mu = {1.010}$, $\sigma = {0.025}$, with a coefficient of 
determination $R^2 =    {0.9993}$ (the fraction of the posterior's shape accounted 
for by the Gaussian fit), confirming the distribution is well described by a normal. 
The posterior peak lies within ${0.374\sigma}$ of unity, confirming that the 
monotonically rising distribution seen in Run~1 was a genuine reflection of the 
data's preference for values near unity, not an artificial truncation effect. 
We also  note 
that the best-fit value i.e., the single parameter combination in the chain that 
minimises $\chi^2$, as distinct from the posterior median or peak, of 
$\varepsilon_{\rm CMB} = {0.9992}$ recovered in Run~1 lies close to the prior wall at 
unity, so the $\chi^2_{\rm min}$ reported for Run~1 in Table~\ref{tab:chi2_bestfit} 
does not represent a global minimum over all parameter space, since the sampler in 
Run~1 cannot explore $\varepsilon_{\rm CMB} > 1$. 
The best-fit $\chi^2_{\rm min}$ for Run~1 should therefore be interpreted cautiously, 
whereas the Run~2 value, where the prior wall is absent, provides the more meaningful 
best-fit comparison with $\Lambda$CDM.  The posterior width $\sigma \approx {0.025}$ 
defines the precision with which Planck 2018
can constrain the photon emissivity, and 
sets the scale below which dust contributions become observationally 
indistinguishable from standard $\Lambda$CDM.

The standard cosmological parameters are recovered with slightly larger uncertainties 
than in Run~1, as expected when $\varepsilon_{\rm CMB}$ is free to explore both sides 
of unity, but their central values are unchanged and in full agreement with the 
baseline (Table~\ref{tab:params_summary}).
The goodness of fit, $\chi^2 = {2795.28^{+6.44}_{-5.28}}$, is again 
indistinguishable from the baseline ($\Delta\chi^2 = {+0.73}$). Now, Run~2 
establishes a consistent picture: the Planck 2018 anisotropy data are compatible with 
$\varepsilon_{\rm CMB} = 1$ to within $\sim 1.5$ per cent at 68 per cent confidence, 
and place a 2$\sigma$ bound of $\varepsilon_{\rm CMB} \gtrsim {0.961}$ (${-3.9}$ per 
cent) at 95 per cent confidence. This is also close to 4.7 per cent lower limit placed by Run~1.
The \citetalias{Gjergo2025} estimate of a 1.4 per cent dust contribution lies 
within this 68 per cent bound and well withing the 95 per cent bounds, 
so widening the prior to allow $\varepsilon_{\rm CMB} > 1$ does not change the basic 
conclusion of Section~\ref{sec:run1}: the data neither detect nor exclude a grey-body 
foreground at the level \citetalias{Gjergo2025} predict. We note that the Run~1 bound ($\varepsilon_{\rm 
CMB} \gtrsim 0.953$, $4.7$ per cent) and this Run~2 2$\sigma$ bound
($\varepsilon_{\rm CMB} \gtrsim 0.961$, $3.9$ per cent) answer slightly different 
questions,  a one-sided versus a two-sided prior, rather than being in tension with 
one another. Therefore, we quote both bounds throughout the paper for completeness.

Run~2 also has implications on the upper end of the \citetalias{Gjergo2025} range. 
Under less  conservative assumptions about the ETG number density (average separation 
$\langle d_0\rangle \approx 9\,\rm Mpc$ rather than $15\,\rm Mpc$), the authors note 
that the dust emission scenario could in principle approach the full CMB energy 
density. The Run~2 constraint $\varepsilon_{\rm CMB} = {1.0104^{+0.025}_{-0.024}}$ 
rules this out at very high significance: a dust contribution approaching or 
exceeding the full CMB energy density would require $\varepsilon_{\rm CMB} \lesssim 
0.01$, excluded at ${42\sigma}$.\footnote{Coincidentally, the significance at which a fully non-primordial
CMB is excluded is $42\sigma$, the Answer to the Ultimate Question of Life,
the Universe, and Everything according to Adams 1979 \citep{Adams1979}.}
Current CMB precision cosmology therefore 
places a strong observational constraint on the upper end of the dust contribution 
even while being unable to detect or rule out the conservative lower bound of 1.4 per cent, estimated by \citetalias{Gjergo2025}.

\subsection{Sensitivity to the transition redshift}
\label{sec:zdep}

In the analyses done above, the transition redshift $z_t = 17$ is held fixed at the
centre of the \citetalias{Gjergo2025} ETG formation epoch. 
To assess whether the constraints on $\varepsilon_{\rm CMB}$ are sensitive to this 
choice, and whether the data carry any information about the epoch of 
$\varepsilon_{\rm CMB}$  transition, we perform a companion run in which $z_t$ is 
treated as a free parameter (hereon referred to as Run~3), 
sampled simultaneously along with $\varepsilon_{\rm CMB} \in [0.5, 1.5]$ and the six 
standard $\Lambda$CDM parameters over the range $z_t \in [5, 50]$. This redshift range
spans from well after hydrogen reionisation $z\sim 6$ \citep{Fan2006,Robertson2015}
to deeply into the matter-dominated era, which would include all physically 
plausible epochs for the ETG starburst population and well beyond.

\begin{figure*}
    \centering
    \includegraphics[width=\textwidth]{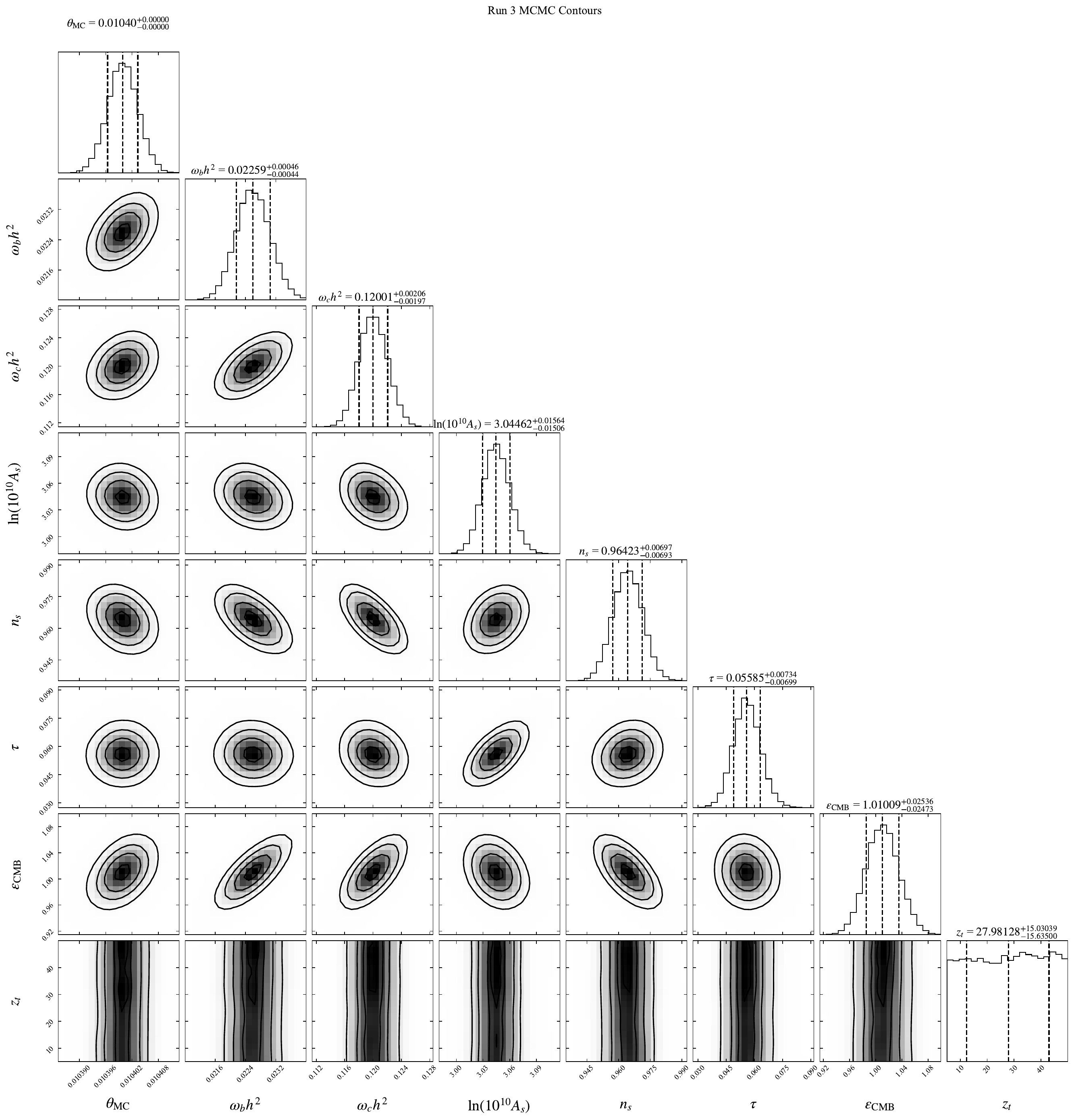}
    \caption{Testing sensitivity to the assumed transition epoch: marginalised posterior distributions for the $\Lambda$CDM + $\varepsilon_{\rm CMB}$ model with both the emissivity $\varepsilon_{\rm CMB}\in[0.5,1.5]$ \emph{and} the transition redshift $z_t \in [5, 50]$ left free (Run~3), constrained using the Planck 2018 TT, TE, EE + low-$\ell$ + BAO + lensing likelihoods. The diagonal panels show the one-dimensional marginalised posteriors of the six standard $\Lambda$CDM parameters together with $\varepsilon_{\rm CMB}$ and $z_t$; the off-diagonal panels show the corresponding two-dimensional joint posteriors, with contours enclosing the 68 and 95 per cent credible regions. Two results stand out. First, the $\varepsilon_{\rm CMB}$-$z_t$ panel shows no significant correlation between the two parameters, so the constraint on $\varepsilon_{\rm CMB}$ obtained here (essentially identical to the fixed-$z_t$ value from Run~2) does not depend on which epoch the transition is assumed to occur at. Second, the one-dimensional posterior of $z_t$ (bottom-right panel, shown enlarged in Fig.~\ref{fig:posterior_zdep}) is entirely flat across the full prior range $[5,50]$, indicating that the Planck 2018 anisotropy data carry no information at all about \emph{when} such a transition might have occurred, only about its amplitude.}
    \label{fig:corner_zdep}
\end{figure*}

\begin{figure*}
    \centering
    \includegraphics[width=\textwidth]{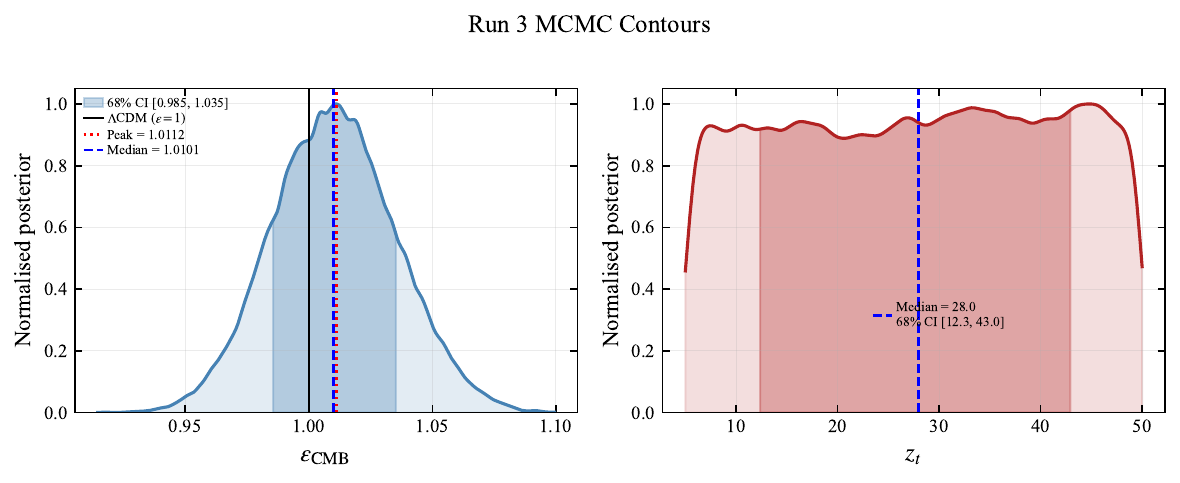}
    \caption{The two one-dimensional posteriors extracted from Run~3 (both $\varepsilon_{\rm CMB} \in [0.5,1.5]$ and $z_t \in [5,50]$ free; TT, TE, EE + low-$\ell$ + BAO + lensing likelihoods), presented side by side to contrast the amplitude and epoch constraints directly. \textit{Left (blue):} the marginalised posterior of $\varepsilon_{\rm CMB}$ is a well-resolved, closed distribution with median $1.0101$ and 68 per cent credible interval $[0.985, 1.036]$; the solid black line marks the standard $\Lambda$CDM value $\varepsilon_{\rm CMB} = 1$ and the blue dashed line marks the posterior median. This constraint is essentially unchanged from the fixed-$z_t$ Run~2 result (Fig.~\ref{fig:posterior_run2}). \textit{Right (red):} in stark contrast, the marginalised posterior of the transition redshift $z_t$ is entirely flat across the full prior range $[5, 50]$ (median $z_t = 27.98$, 68 per cent credible interval $[12.27, 43.08]$, consistent with a uniform distribution), showing that the Planck 2018 anisotropy data constrain how large a dust contribution can be, is excluded at highbut are completely insensitive to when it occurred, for any transition within the sampled range.} significance ($42\sigma$).
    \label{fig:posterior_zdep}
\end{figure*}

The marginalised posteriors of $\varepsilon_{\rm CMB}$ and $z_t$ are shown in 
Fig.~\ref{fig:posterior_zdep}, with the full joint posterior in 
Fig.~\ref{fig:corner_zdep}. 
The result for $z_t$ is interesting: the posterior is
entirely flat across the prior range, with a median of $z_t =27.98^{+15.10}_{-15.71}$
fully consistent with a uniform distribution. This shows that the Planck 2018 
anisotropy data has essentially no sensitivity to the epoch of $\varepsilon_{\rm CMB}$
transition for any $z_t \in [5, 50]$.

This insensitivity is physically transparent. All values of $z_t$  within this range 
lie well below the epoch of recombination  ($z \approx 1090$), so the modification to 
$\rho_\gamma$ active at  $z > z_t$ always encompasses the recombination epoch 
regardless of the precise value of $z_t$. The CMB anisotropy spectrum is sensitive to 
the \textit{integrated} effect of the radiation content on the sound horizon and the 
acoustic peak structure, accumulated over the full expansion history between the 
$z_t$ and the present day. 
Because every value of $z_t$ explored here is well below the recombination redshift 
$z \approx 1090$, the modified photon energy density is already active throughout 
recombination and the subsequent acoustic-peak formation for any $z_t$ in this range: 
shifting $z_t$ earlier or later within $[5,50]$ does not change what the anisotropy 
spectrum ``sees'', since recombination physics is identically affected in either 
case. This is the specific reason the data constrain the \textit{amplitude} of the 
$\varepsilon_{\rm CMB}$ but not its \textit{epoch} ($z_t$) over this range.
The picture could differ if $z_t$ is pushed above $z \approx 1090$, since a 
transition occurring \textit{after} recombination would leave the acoustic peak 
structure itself unmodified regardless of $\varepsilon_{\rm CMB}$, a regime not 
explored in the current analysis. 

This insensitivity to $z_t$ has an important implication for the robustness
of the constraint. Our measurement of $\varepsilon_{\rm CMB}$ (in Run~2) and
the lower bounds on it (in Run~1) are not specific to the fiducial choice
$z_t = 17$: they apply equally to any CMB foreground established well after
recombination, regardless of the precise epoch of ETG formation.
The constraint on $\varepsilon_{\rm CMB}$ in Run~3, 
$\varepsilon_{\rm CMB} = {1.0101^{+0.0255}_{-0.025}}$, is essentially identical to 
that obtained with $z_t$ fixed, confirming that the result of Section~\ref{sec:run1} and Section~\ref{sec:run2}
is fully robust to the assumed transition redshift. The joint posterior of 
$\varepsilon_{\rm CMB}$ and $z_t$ in Fig.~\ref{fig:corner_zdep} shows no significant
correlation between the two parameters, further supporting this interpretation.

The median $\chi^2 = {2795.37^{+6.52}_{-5.29}}$ in Run~3 is slightly elevated relative to the
fixed-$z_t$ runs, with $\Delta\chi^2 \approx {+0.82}$ compared to the baseline. This confirms that adding $z_t$ as a free parameter introduces no meaningful
degradation of fit, consistent with the flat $z_t$ posterior: when a parameter is
entirely unconstrained by the data, the likelihood is insensitive to its value and
the sampler explores the full prior volume without penalty. The goodness-of-fit of the
best-sampled points in Run~3 is indistinguishable from the baseline, confirming that
the result reflects sampling geometry rather than any genuine model misfit. 
However, it should not be interpreted as evidence against the presence of 
$\varepsilon_{\rm CMB}$ that is not unity.

The flat $z_t$ posterior also has implications for future observations. The inability
of Planck 2018 to distinguish between $\varepsilon_{\rm CMB}$ transitions 
at $z_t = 5$ and $z_t = 50$ points to a genuine physical degeneracy rather than a 
limitation of the data quality alone. Breaking this degeneracy would require 
exploiting scale-dependent
signatures in the power spectrum-such as the Silk damping tail at small angular
scales, which is sensitive to the detailed history of the radiation content near
recombination-rather than just the integrated peak structure probed at the angular
scales accessible to Planck. Next-generation experiments such as the Simons 
Observatory \citep{Ade2019SimonsObservatory} and CMB-S4 \citep{Abazajian2016CMBS4}, 
with their substantially improved sensitivity at high $\ell$, may be able to 
constrain $z_t$ or $\varepsilon_{\rm CMB}$ directly, providing a new window on the 
epoch of ETG formation or the possibility of a non-primordial CMB radiation model 
through the CMB power spectrum.

\begin{figure}[ht!]
    \centering
    \includegraphics[width=\textwidth]{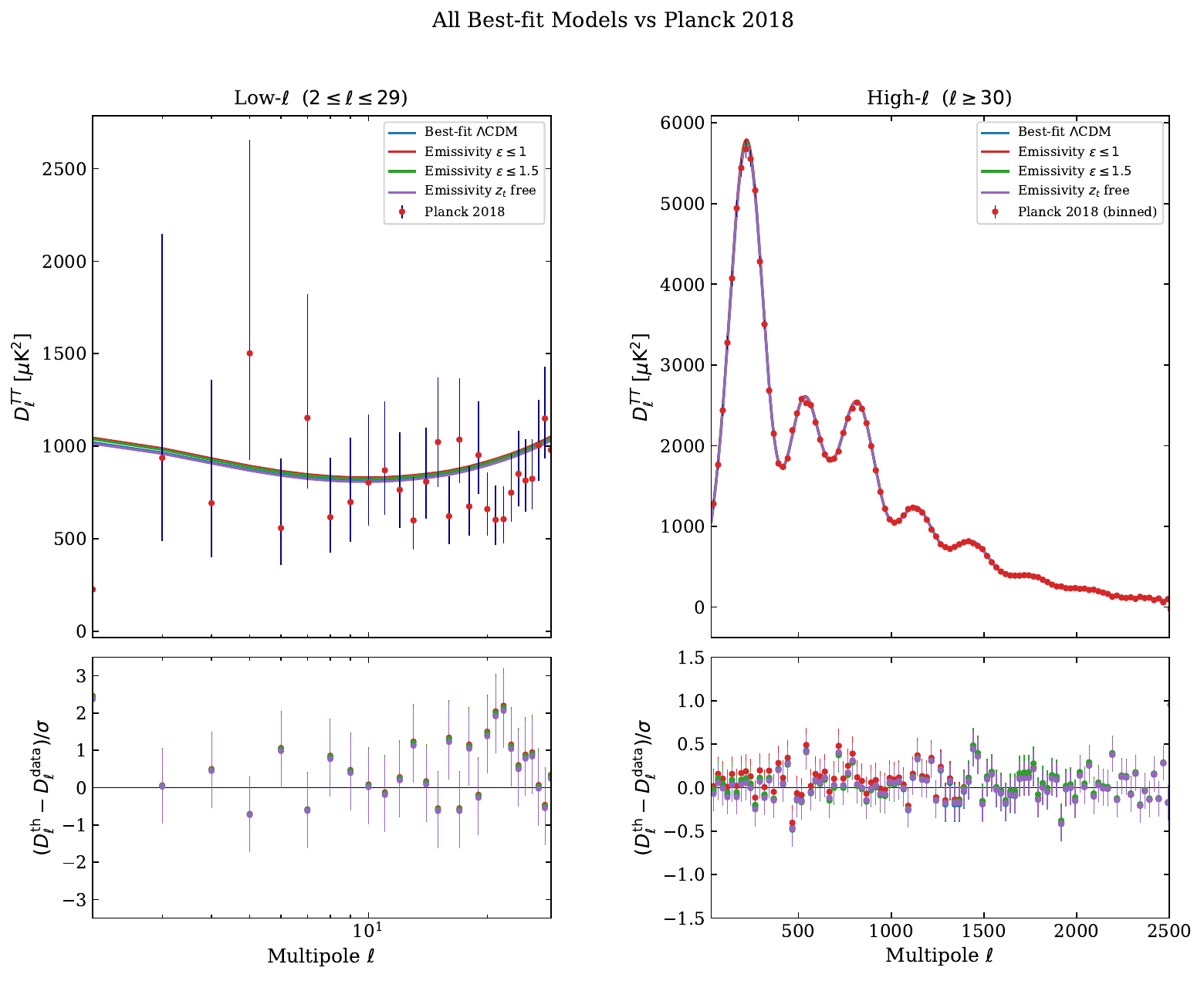}
    \caption{Visual goodness-of-fit check: best-fit CMB TT angular power spectra, $D_\ell^{\rm TT} = \ell(\ell+1)C_\ell^{\rm TT}/2\pi$, for all four MCMC runs overlaid on the Planck 2018 band powers (black points with error bars), split into the cosmic-variance-dominated low-$\ell$ regime ($2 \leq \ell \leq 29$, left panels, log multipole scale) and the acoustic-peak high-$\ell$ regime ($\ell \geq 30$, right panels, linear scale). The top panels show the spectra themselves and the bottom panels show the normalised residuals, $(D_\ell^{\rm th} - D_\ell^{\rm data})/\sigma$, relative to the Planck data. The four best-fit theoretical curves - baseline $\Lambda$CDM (blue), Run~1 with $\varepsilon_{\rm CMB}\leq 1$ (red), Run~2 with $\varepsilon_{\rm CMB}\leq 1.5$ (green), and Run~3 with $z_t$ free (purple) - lie essentially on top of one another and are visually indistinguishable across the full multipole range, with residuals consistent with zero throughout. This confirms, independently of the posterior and $\chi^2$ statistics reported in the text, that none of the grey-body extensions alters the shape of the acoustic peak structure by an amount detectable at the resolution of this plot, consistent with the negligible $\Delta\chi^2$ values reported in Table~\ref{tab:chi2_bestfit}.}
    \label{fig:ps_combined}
\end{figure}
%
\subsection{CMB TT power spectrum: best-fit models vs Planck 2018}
\label{sec:bestfit_ps}

Before turning to the broader implications of these results, we verify the quality of 
fit visually by comparing the best-fit theoretical CMB TT power spectra against the 
Planck 2018 binned band powers. For each run, the best-fit parameter vector is 
identified as the chain sample minimising $\chi^2$(Table~\ref{tab:chi2_bestfit}), 
and the corresponding power 
spectrum is computed using \textsc{camb} with the emissivity modification active. The 
Planck 2018 TT band powers and their asymmetric error bars are taken from
the public data release (version R3.01), available via the Planck Legacy Archive, 
\url{https://pla.esac.esa.int/}.

Fig.~\ref{fig:ps_combined} shows all four best-fit models overlaid on the Planck 2018 
data, split into the low-$\ell$ ($2 \leq \ell \leq 29$) and high-$\ell$ ($\ell \geq 
30$) regimes. The four curves are visually barely distinguishable at the resolution 
of this figure, consistent with the small $\Delta\chi^2$ values reported in 
Table~\ref{tab:chi2_bestfit}. The combined residuals panels of 
Fig.~\ref{fig:ps_combined} confirm this: all four models produce normalised residuals 
$(D_\ell^{\rm th} - D_\ell^{\rm data})/\sigma$ that are consistent with zero across 
the full multipole range $30 \leq \ell \leq 2500$, with no systematic trend 
distinguishing the $\varepsilon_{\rm CMB}$ runs from the baseline $\Lambda$CDM fit.
The low-$\ell$ panel of Fig.~\ref{fig:ps_combined} likewise shows no differentiation 
between the four models in the cosmic-variance-dominated regime. The low-$\ell$ 
scatter in this panel reflects cosmic variance, while the high-$\ell$ residuals are 
shown as binned means and are consistent with zero throughout the multipole range 
probed.

\section{Discussion}
\label{sec:discussion}

\subsection{Short summary of the results}

We have presented the first constraints on a non-primordial contribution to
the CMB using the full Planck 2018 CMB anisotropy data set in
combination with lensing and BAO measurements, motivated by the work of 
\citetalias{Gjergo2025} suggesting that dust emission from  
dust-enshrouded formation of massive ETGs at $z \sim 15$-$20$ can contribute 
significantly to the observed energy density of the CMB. 
We parametrise the dust contribution through a single dimensionless emissivity
parameter $\varepsilon_{\rm CMB}$, which rescales the standard primordial photon 
energy density above a transition redshift $z_t$ (equation~\ref{eq:rho_eff}). This
modification is implemented self-consistently in \textsc{camb}, as discussed in Section~\ref{sec:model}. A standard six-parameter flat $\Lambda$CDM baseline run confirms
that the pipeline recovers all Planck 2018 parameter values with no systematic 
offsets.

Our main results are summarized as follows. In Run~1 described in 
Section~\ref{sec:run1} ($\varepsilon_{\rm CMB} \in [0.5, 1.0]$, fixed $z_t = 17$), 
the posterior yields a median of $\varepsilon_{\rm CMB} = {0.987}$ with a 68 per cent 
credible interval of ${[0.9713, 0.9963]}$ and a 95 per cent lower bound of 
$\varepsilon_{\rm CMB} \gtrsim {0.953}$, allowing a dust contribution of up to $4.7 
{5}$ per cent. In Run~2 mentioned in Section~\ref{sec:run2} ($\varepsilon_{\rm CMB} 
\in [0.5, 1.5]$), the posterior is a well-resolved Gaussian measuring
$\varepsilon_{\rm CMB} = {1.0104^{+0.025}_{-0.024}}$. This 
confirms that the rising posterior in Section~\ref{sec:run1} reflects the data's 
genuine preference for $\varepsilon_{\rm CMB}$ 
values near unity rather than a prior truncation effect. In a third run with both
$\varepsilon_{\rm CMB}$ and $z_t$ free (Section~\ref{sec:zdep}), 
the $\varepsilon_{\rm CMB}$ constraint is unchanged while the posterior of 
$z_t$ is entirely flat across $5 \lesssim z_t \lesssim 50$, demonstrating that the 
constraint on $\varepsilon_{\rm CMB}$ is robust to the assumed
transition epoch. The $\chi^2$ values across all four runs are summarised in
Table~\ref{tab:chi2_summary}, and are indistinguishable within their uncertainties.


\begin{table}
\centering
\caption{Summary of goodness-of-fit statistics for all four MCMC runs. The $\chi^2$ values are the median of the chain posterior distribution with 68 per cent credible intervals; the median is reported here to characterise the typical goodness-of-fit across the full chain, as distinct from the best-fit minimum reported in Table~\ref{tab:chi2_bestfit}. $\Delta\chi^2$ is the difference in the posterior median relative to the posterior median of the baseline $\Lambda$CDM run.}
\label{tab:chi2_summary}
\begin{tabular}{lccc}
\hline
Run & $\varepsilon_{\rm CMB}$ prior & $\chi^2$ & $\Delta\chi^2$ \\
\hline
Baseline $\Lambda$CDM &-& ${2794.55^{+6.28}_{-5.14}}$ &-\\
Run 1 (Grey-Body)      & $[0.5, 1.0]$ & ${2795.72^{+6.50}_{-5.29}}$ & ${+1.17}$ \\
Run 2 (Super Black-Body)    & $[0.5, 1.5]$ & ${2795.28^{+6.44}_{-5.28}}$ & ${+0.73}$ \\
Run 3 ($z_t$ free)    & $[0.5, 1.5]$ & ${2795.37^{+6.51}_{-5.29}}$ & ${+0.82}$ \\
\hline
\end{tabular}
\end{table}
\begin{table}
\centering
\caption{Best-fit goodness-of-fit statistics for all four MCMC runs. The $\chi^2_{\rm min}$ values are the minimum $\chi^2$ found across all post-burn-in chain samples. $\Delta\chi^2$ is the difference relative to the baseline $\Lambda$CDM run.}
\label{tab:chi2_bestfit}
\begin{tabular}{lccc}
\hline
Run & $\varepsilon_{\rm CMB}$ prior & $\chi^2_{\rm min}$ & $\Delta\chi^2$ \\
\hline
Baseline $\Lambda$CDM & $-$      & ${2777.02}$ & $-$                \\
Run 1 (Grey-Body)      & $[0.5, 1.0]$ & ${2778.98}$ & ${+1.96}$ \\
Run 2 (Super Black-Body)    & $[0.5, 1.5]$ & ${2777.18}$ & ${+0.15}$ \\
Run 3 ($z_t$ free)    & $[0.5, 1.5]$ & ${2778.67}$ & ${+1.65}$  \\
\hline
\end{tabular}
\end{table}
The best-fit $\chi^2_{\rm min}$ values, extracted as the minimum across all 
post-burn-in chain samples, are summarised in Table~\ref{tab:chi2_bestfit}. 
All $\Delta\chi^2_{\rm min}$ values are positive and small, confirming that 
no extended model provides a meaningfully better fit than standard $\Lambda$CDM 
at the best-fit level. 

The standard $\Lambda$CDM cosmological parameters are summarised across all runs in
Table~\ref{tab:params_summary}. The shifts between the baseline and the grey-body runs
are small and statistically insignificant throughout, with the exception of $\omega_b h^2$
in Run~1, which shifts by $\sim 1.1\sigma$ relative to its own posterior uncertainty; no
other parameter shifts by more than $\sim 1\sigma$ in any run.
The angular scale $\theta_{\rm MC}$ is kept entirely stable
across all runs.

\begin{table*}
\centering
\caption{Marginalised posterior constraints (median with 68 per cent credible intervals) for the standard $\Lambda$CDM parameters across all four runs, compared against the Planck 2018 TT,TE,EE+lowE+lensing+BAO results \citep{PlanckCollaboration2020}. All values are consistent with Planck within their respective uncertainties.}
\label{tab:params_summary}
\renewcommand{\arraystretch}{1.6}
\resizebox{\textwidth}{!}{%
\begin{tabular}{lcccccc}
\hline
Parameter & Planck 2018 & Baseline $\Lambda$CDM & Run 1 & Run 2 & Run 3 ($z_t$ free) \\
\hline
$100\theta_{\rm MC}$ &
$1.04101 \pm 0.00029$ &
$1.03990^{+0.00030}_{-0.00030}$ &
$1.03980^{+0.00030}_{-0.00030}$ &
$1.04000^{+0.00030}_{-0.00040}$ &
$1.04000^{+0.00030}_{-0.00040}$ \\
$\omega_b h^2$ &
${0.02242 \pm 0.00014}$ &
${0.02243^{+0.00013}_{-0.00013}}$ &
${0.02218^{+0.00022}_{-0.00028}}$ &
${0.02261^{+0.00045}_{-0.00045}}$ &
${0.02260^{+0.00046}_{-0.00044}}$ \\
$\omega_c h^2$ &
${0.11933 \pm 0.00091}$ &
${0.11932^{+0.00094}_{-0.00093}}$ &
${0.11824^{+0.00120}_{-0.00134}}$ &
${0.12005^{+0.00199}_{-0.00196}}$ &
${0.12001^{+0.00207}_{-0.00198}}$ \\
$\ln(10^{10} A_s)$ &
${3.047 \pm 0.014}$ &
${3.04688^{+0.01446}_{-0.01401}}$ &
${3.05063^{+0.01484}_{-0.01426}}$ &
${3.04461^{+0.01519}_{-0.01466}}$ &
${3.04462^{+0.01572}_{-0.01515}}$ \\
$n_s$ &
${0.9665 \pm 0.0038}$ &
${0.96657^{+0.00366}_{-0.00366}}$ &
${0.97027^{+0.00503}_{-0.00446}}$ &
${0.96427^{+0.00689}_{-0.00692}}$ &
${0.96423^{+0.00700}_{-0.00697}}$ \\
$\tau$ &
${0.0561 \pm 0.0071}$ &
${0.05606^{+0.00732}_{-0.00689}}$ &
${0.05649^{+0.00746}_{-0.00695}}$ &
${0.05594^{+0.00723}_{-0.00691}}$ &
${0.05585^{+0.00737}_{-0.00703}}$ \\
$H_0\,[\mathrm{km\,s^{-1}\,Mpc^{-1}}]$ &
${67.66 \pm 0.42}$ &
${67.64^{+0.40}_{-0.40}}$ &
${67.49^{+0.44}_{-0.41}}$ &
${67.76^{+0.50}_{-0.53}}$ &
${67.75^{+0.49}_{-0.46}}$ \\
$\sigma_8$ &
${0.8102 \pm 0.0060}$ &
${0.8104^{+0.0058}_{-0.0061}}$ &
${0.8136^{+0.0070}_{-0.0063}}$ &
${0.8084^{+0.0078}_{-0.0077}}$ &
${0.8085^{+0.0078}_{-0.0079}}$ \\
\hline
\\[-1.3em]
$\varepsilon_{\rm CMB}$ & - & - &
${0.98674^{+0.00952}_{-0.01544}}$ &
${1.01038^{+0.02497}_{-0.02442}}$ &
${1.01009^{+0.02548}_{-0.02487}}$ \\
$z_t$ & - & - & $17$ (fixed) & $17$ (fixed) &
${27.98^{+15.03}_{-15.63}}$ \\
\hline
\end{tabular}%
}
\end{table*}

\subsection{Implications for CMB parameter inference}
\label{sec:param_implications}

A recurring concern in precision cosmology is whether unmodelled foreground components
can bias the inferred values of cosmological parameters. Our results address this
concern. None of the parameter shifts
between the baseline and Run~1 is statistically significant, and none would alter any
cosmological conclusion drawn from the Planck 2018 analysis. The shifts in $\omega_b
h^2$ and $\omega_c h^2$ are in the direction expected from the partial degeneracy with
the photon energy density discussed in Section~\ref{sec:run1}, but their magnitudes
are well within the existing Planck uncertainties.

The derived Hubble constant $H_0$ and $\sigma_8$ 
are of particular interest in the context of the 
Hubble and $\sigma_8$ tensions. 
Because $\theta_{\rm MC}$ is the directly sampled parameter in 
\texttt{cobaya}, and because our \texttt{camb} modification updates the 
\texttt{dsound\_da\_approx}$^\star$ function that converts $\theta_{\rm MC}$ to 
$H_0$, the derived Hubble constant adjusts self-consistently with the modified 
radiation content. 
The baseline $\Lambda$CDM run gives $H_0 = 67.64^{+0.40}_{-0.40}\,\mathrm{km\,s^{-1}\,Mpc^{-1}}$, 
in excellent agreement with Planck~2018 ($67.66\pm0.42$). 
Run~1 gives $H_0 = 67.49^{+0.44}_{-0.41}$, a shift of only
$\sim 0.37\sigma$ relative to the baseline (in units of the run's own uncertainty); 
Run~2 gives
$H_0 = 67.76^{+0.50}_{-0.53}$, a shift of $\sim 0.24\sigma$; Run~3 gives
$H_0 = 67.75^{+0.49}_{-0.46}$, $\sim 0.21\sigma$. 
Similarly, $\sigma_8 = 0.8104^{+0.0058}_{-0.0061}$
in the baseline shifts to $0.8136^{+0.0070}_{-0.0063}$ (Run~1, $\sim 0.46\sigma$) and
$0.8084^{+0.0078}_{-0.0077}$ (Run~2, $\sim 0.26\sigma$). 
In every case the shift in $H_0$ and $\sigma_8$ is well under
$1\sigma$, confirming that the $\varepsilon_{\rm CMB}$ modification at the 
\citetalias{Gjergo2025} level neither alleviates nor exacerbates the Hubble tension, 
and leaves 
$\sigma_8$ correspondingly unaffected. A larger dust contribution, approaching
a higher value of $H_0$, could in principle produce more significant parameter shifts, which might result in lowering the Hubble tension
but such a contribution is strongly disfavoured by the data (Section~\ref{sec:run2}).

Our results can be read in two directions. For precision cosmology, they provide a 
quantitative robustness statement: the Planck 2018 parameter inference is stable 
against an entire class of unmodelled, late-established isotropic foregrounds at the 
few-per-cent level, extending the standard foreground-marginalisation arguments to a 
component that modifies the background radiation density itself. For non-standard 
cosmology, they demonstrate that proposals for a partially non-primordial CMB need 
not be dismissed a priori but can be confronted quantitatively with anisotropy data: 
the \citetalias{Gjergo2025} scenario survives this confrontation at its conservative amplitude, while 
its optimistic extreme does not. 

We also emphasise that leaving \textsc{recfast} \citep{Seager2000} unmodified in our
implementation is a deliberate modelling choice, not a claim that the photon
temperature is unaffected by the emissivity rescaling in a fully self-consistent
treatment. Because $\rho_\gamma \propto T_\gamma^4$ for a true blackbody, one might be
tempted to infer an effective temperature 
$T_\gamma^{\rm eff}(z) = \varepsilon_{\rm CMB}^{1/4}\,T_{\rm 0}(1+z)$ from the 
rescaled energy density entering the background and perturbation equations. 
However, the grey-body deficit we model is assumed to remain Planckian in form, 
so no unique effective temperature can be assigned to it via the Stefan-Boltzmann 
relation alone.
We instead treat the emissivity rescaling as acting on the integrated radiation energy
density that enters the background expansion, the Boltzmann hierarchy, and the sound
horizon, while leaving the photon occupation numbers, and hence the Saha equation and
atomic transition rates entering \textsc{recfast}, at their standard values.

The $\varepsilon_{\rm CMB}$ modification can also be expressed as an effective shift 
in the number of relativistic species $N_{\rm eff}$, providing a useful point of
comparison with the broader $N_{\rm eff}$ and big bang nucleosynthesis literature
\citep{Cyburt2016BBN}. In our implementation, only the photon energy density is
rescaled, while the neutrino energy density remains unchanged, so that the total
radiation density becomes
\begin{equation}
    \rho_r =
    \varepsilon_{\rm CMB}\rho_\gamma +
    \rho_\nu
    =
    \rho_\gamma\left[
    \varepsilon_{\rm CMB}
    +
    \frac78\left(\frac{4}{11}\right)^{4/3}N_{\rm eff}
    \right].
\end{equation}
Requiring this modified radiation density to be reproduced by the standard
expression with an effective shift $\Delta N_{\rm eff}$ gives
\begin{equation}
    \Delta N_{\rm eff}
    =
    \frac{\varepsilon_{\rm CMB}-1}
    {\frac78\left(\frac{4}{11}\right)^{4/3}}
    \approx
    4.4\,(\varepsilon_{\rm CMB}-1),
    \label{eq:dneff}
\end{equation}
where the numerical coefficient is evaluated for the standard neutrino-photon
temperature ratio. The Run~2 central value,
$\varepsilon_{\rm CMB}=1.0103$, corresponds to
$\Delta N_{\rm eff}\approx +0.045$, while the
\citetalias{Gjergo2025} estimate
$\varepsilon_{\rm CMB}\approx0.986$ gives
$\Delta N_{\rm eff}\approx-0.062$. The $1\sigma$ lower bound from
Run~1, $\varepsilon_{\rm CMB}\gtrsim0.9713$, corresponds to
$\Delta N_{\rm eff}\gtrsim-0.13$. These effective shifts are well within
the current Planck 2018 constraint,
$N_{\rm eff}=2.99\pm0.17$, providing a useful consistency check.

This mapping should not be interpreted as a physical equivalence between the
two parametrisations. A change in $N_{\rm eff}$ modifies the neutrino energy
density while leaving the photon density unchanged, whereas our model rescales
only the photon energy density. Consequently, the two models differ in their
perturbation evolution and would make different predictions for observables
such as the damping tail, acoustic phase shifts, and spectral distortions.
Equation~(\ref{eq:dneff}) is therefore intended only as a convenient
background-level correspondence for comparison with the broader
$N_{\rm eff}$ literature \citep{Cyburt2016BBN, BashinskySeljak2004, Follin2015, BaumannGreenWallisch2016}.

%

%

\subsection{Limitations}
\label{sec:limitations}

The present analysis has a number of limitations worth discussing openly.

The step-function parametrisation of the dust contribution at $z_t$ is a minimal
phenomenological choice, and a fairly crude one. In reality, the dust emission from
ETG formation would have built up over a finite epoch ($15 \lesssim z \lesssim 20$) rather
than switching on instantaneously, so a smooth emissivity profile tracking the ETG
star-formation history would be a more realistic model. We have not attempted this here.
That said, since the data are shown in Section~\ref{sec:zdep} to be insensitive to the
transition epoch even under the step-function approximation, we doubt a smooth profile
would change our conclusions in any qualitative way, though this remains to be checked
explicitly in future work.

We have also assumed throughout that the dust foreground contributes only to the
homogeneous photon energy density, with no anisotropic component, on the grounds that
emission distributed across the observable Universe at $z \sim 15$-$20$ should be largely
isotropic on Planck angular scales. This assumption may not hold exactly. \citetalias{Gjergo2025}
point out that the southern celestial hemisphere contains a significantly larger number of
early-type galaxies than the north, raising the possibility of a hemispherical asymmetry in
the dust emission - one that could even connect to the known CMB hemispherical power
asymmetry, where the southern hemisphere shows more large-scale power than the northern one
\citep{PlanckCollaboration2020}. We note that the complementary monopole analysis of
\citep{Corredoira2026_resnote}, which fits the COBE-FIRAS spectrum with a single global dust
component parametrised by an emissivity index $\beta$ and a Gaussian spread $\sigma_d$ in
progenitor formation redshift, similarly treats the dust contribution as spatially uniform
and makes no attempt to model a hemispherical or otherwise anisotropic distribution of the
emitting sources. Modelling a spatially modulated grey-body component therefore remains an
open question for both the monopole and anisotropy tests of the scenario, and is beyond
what we attempt here.

\subsection{Implications of the non-primordial origin of the CMB} \label{sec:implications}

\subsubsection{The CMB monopole: a complementary and currently
tighter constraint}

The CMB monopole spectrum offers a test of the dust scenario that is
complementary to the anisotropy analysis performed here, and recent
work indicates it is presently the more restrictive of the two for
realistic dust properties. COBE-FIRAS constrains departures from a
perfect blackbody to $\Delta I/I \lesssim 10^{-5}$ across
60--630\,GHz \cite{Fixsen1996}. Following the appearance of \citetalias{Gjergo2025},
\cite{Corredoira2026_resnote} fitted the FIRAS monopole with a
superposition of a primordial blackbody and redshifted dust emission
from $z_t = 17$, parametrised by the dust emissivity index $\beta$
and the dispersion $\sigma_d$ of progenitor formation redshifts. For
the physically realistic region $\beta \geq 0.5$ and
$\sigma_d \geq 0.8$ (the latter matching the width of the \citetalias{Gjergo2025} ETG
birth function), the dust contribution to the FIRAS flux is limited
to $<1.3$ per cent at 95 per cent confidence, three times tighter
than our anisotropy bound of $\sim$ 4-5 per cent, and marginally
below the conservative \citetalias{Gjergo2025} estimate of 1.4 per cent. For realistic
dust, therefore, the monopole spectral shape, rather than the
anisotropy power spectrum, is the limiting observational constraint
on the scenario at present.

The two probes rest on different physics and different systematics,
and neither subsumes the other. Our anisotropy test is achromatic by
construction (Section~\ref{sec:param}):
$\varepsilon_{\rm CMB}$ rescales the background photon energy density
and constrains its dynamical imprint on the expansion history, sound
horizon, and acoustic peak structure, irrespective of the spectral
shape of the non-primordial component. In the decomposition of
\cite{Ellis2013}, who used FIRAS to bound an achromatic grey-body
deviation arising from distance-duality violation,
our parametrisation probes the achromatic sector at the level of the energy density, whereas the
monopole fits of \cite{Corredoira2026_resnote,Ellis2013} are driven by
the chromatic residual. The monopole bound consequently depends on
assumptions about $\beta$, the dust temperature distribution, and the
FIRAS Galactic foreground subtraction (which
\cite{Corredoira2026_resnote} show relaxes the bound from 1.0 to 1.3 per
cent); the anisotropy bound depends on none of these.

This distinction becomes decisive in the spectrally degenerate limit.
For Planckian dust emission ($\beta = 0$) from a single formation
epoch ($\sigma_d = 0$), \cite{Corredoira2026_resnote} find the FIRAS
monopole compatible with 100\% dust emission i.e., a microwave background
of \emph{entirely} dust origin cannot be excluded by its spectrum,
because the redshifted dust and primordial components are spectrally
identical. It is precisely here that our anisotropy constraint takes
over. A fully non-primordial CMB corresponds to
$\varepsilon_{\rm CMB} \approx 0$, which destroys the acoustic peak
structure regardless of how closely the dust spectrum mimics a
blackbody; our Run~2 measurement excludes this at $\sim 42\sigma$
(Section~\ref{sec:run2}). The long-standing class of proposals in which the CMB is wholly thermalised
starlight or dust emission \citep[and references
therein]{Wesson1975The,Wickramasinghe1975A,AguirreHaiman2000,LopezCorredoira2013NonStandard,Cirkovic2017AltCMB}
survives the FIRAS constraint only by retreating into this degenerate limit;
the anisotropy power spectrum closes that loophole. The two
probes are thus complementary in a precise sense: the monopole is
tighter wherever the dust is chromatic, while the anisotropy spectrum
excludes the achromatic extreme that the monopole cannot touch.

The physical reason a per-cent-level contamination survives FIRAS at
all is the temperature near-coincidence noted in
Section~\ref{sec:run1}, assuming that this is not hard-coded in the construction
of the \citetalias{Gjergo2025} model.
Energy conservation requires
\begin{equation}
(1-\varepsilon_{\rm CMB})\,T_{\rm dust}^4
 + \varepsilon_{\rm CMB}\,T_{\rm prim}^4 = T_0^4 \, ;
\label{eq:energycons}
\end{equation}
for $1-\varepsilon_{\rm CMB} = 0.014$ and
$T_{\rm dust} \approx 2.78$\,K this gives
$T_{\rm prim} \approx 2.72$\,K, so the two components differ by only
$\Delta T \approx 0.06$\,K. Treating both as blackbodies, their
superposition with mixing fraction $f = 1-\varepsilon_{\rm CMB}$
produces a leading-order $y$-type distortion of effective amplitude
\begin{equation}
y_{\rm eff} \sim \tfrac{1}{2}\, f\,(1-f)
 \left(\frac{\Delta T}{T_0}\right)^{\!2} \approx 3\times 10^{-6},
\label{eq:yeff}
\end{equation}
below the FIRAS limit $|y| < 1.5\times10^{-5}$. It is the
chromaticity of realistic dust emission,
$\epsilon_\nu \propto \nu^{\beta}$ with $\beta \simeq 1.5$--$2$,
together with the spread of formation
redshifts, that breaks the spectral degeneracy and drives the
monopole bound down to the per-cent level. Because the injection
occurs at $z \sim 15$--$20$, far below the thermalisation redshift
$z_{\rm th} \approx 2\times10^{6}$ \citep{ChlubaSunyaev2012,KhatriSunyaev2012},
this residual distortion cannot be erased and survives to the
present epoch \citep{ChlubaKhatriSunyaev2012,Chluba2016}.

Future spectral measurements will sharpen this test decisively. The
tightened FIRAS-based $y$-distortion bound of \citep{Fabbian2025}
already approaches the $y_{\rm eff}$ estimated above, and a
PIXIE-class mission \citep{Kogut2011,Kogut2025PIXIE} would improve monopole
sensitivity by several orders of magnitude \citep{Chluba2021SpectralDistortions,
Chluba2019Astro2020},
rendering a 1.4 per cent dust component either detected
or excluded outright.

\subsubsection{Relation to other non-standard CMB scenarios}

\label{sec:nonstandard}

The $\varepsilon_{\rm CMB}$-$z_t$ framework developed here is not restricted 
to the \citetalias{Gjergo2025} dust scenario. A long line of non-standard 
cosmology models has proposed non-primordial or otherwise non-thermal origins 
for some part of the observed CMB, reviewed by \citep{LopezCorredoira2013NonStandard} and by \citep{Cirkovic2017AltCMB}. 
These include tired-light cosmology 
\citep{Zwicky1929TiredLight, Carlos25_CMB_tired_light}, $R_{h} = ct$ Universe \cite{Melia2018RhCTReview, Melia2022A, Fujii20_CMB_Rh_ct},  
quasi-steady-state cosmology (QSSC) variants 
\citep{HoyleBurbidgeNarlikar1993QSSC, Narlikar2002Inhomogeneities}, and more 
recent covarying-coupling-constants proposals motivated by the same JWST early-
galaxy population \citep{Gupta2023CCCTL}. Notably, \citep{Gupta2023CCCTL}
explicitly flags CMB isotropy as a difficulty for tired-light models without 
providing a quantitative test against the anisotropy power spectrum. Because our 
framework is defined purely in terms of an energy-density contribution above a 
transition redshift, without reference to any specific emission mechanism, it is agnostic to the physical origin of the
non-standard component: the same two-parameter test applied here to the
\citetalias{Gjergo2025} scenario could equally be applied to the achromatic grey-
body signature predicted by distance-duality violation \citep{Ellis2013, Chiara26_distance_duality_DESI} or to the non-thermal contributions proposed in 
these alternate cosmologies, using the identical Planck~2018 data set
employed here. 
We leave a dedicated application of the framework to these scenarios for future
work, but note that the method itself provides a general and reusable 
observational test for any proposal in which some fraction of the present-day 
CMB is non-primordial.

\subsection{Future directions}
\label{sec:future}

The most immediate extension of this work is to use newer data.
Improved small-scale measurements are now available from both
ACT~DR6 \citep{Louis2025ACTDR6, Calabrese2025ACTDR6Extended,
Madhavacheril2024ACTDR6Lensing, Qu2024ACTDR6LensingPowerSpectrum,
Beringue2025ACTDR6Foregrounds} and SPT-3G \citep{SPT3G2025}, both of
which reach smaller angular scales with better polarisation
sensitivity than \textit{Planck}. Because the Silk damping tail
responds more strongly to the photon energy density at recombination
than the intermediate scales that dominate the \textit{Planck}
constraint, applying the $\varepsilon_{\rm CMB}$--$z_t$ framework to
these data sets should tighten the bound on $\varepsilon_{\rm CMB}$
directly, and a joint ACT\,+\,SPT\,+\,\textit{Planck} analysis may in
addition begin to break the epoch degeneracy identified in
Section~\ref{sec:zdep}. Replacing the BAO compilation used here with
DESI~DR2 \citep{AbdulKarim2025DESIDR2} is a further straightforward
improvement.

On the modelling side, two refinements suggest themselves. First, the
step function of equation~(\ref{eq:rho_eff}) could be replaced by a
smooth emissivity profile tracking the ETG star-formation history;
given the insensitivity of the data to the transition epoch
(Section~\ref{sec:zdep}), we expect this to matter only once
higher-precision data are available, but it should be verified
explicitly. Second, the isotropic treatment adopted here could be
generalised to a spatially modulated dust component, motivated by the
hemispherical asymmetry in the ETG distribution noted in
Section~\ref{sec:limitations}, allowing a direct test of whether the
dust foreground connects to the known large-scale power asymmetry of
the CMB.

\section{Summary}
\label{sec:conclusions}

\begin{enumerate}

\item Combining Planck 2018 CMB anisotropy data with lensing and BAO measurements in Run~1 ($\varepsilon_{\rm CMB}<1$) gives a 95 per cent lower bound of $\varepsilon_{\rm CMB} \gtrsim {0.953}$ on the primordial photon emissivity above $z_t = 17$ (Figs.~\ref{fig:corner_run1} and~\ref{fig:posterior_run1}), so a dust contribution of up to $\sim {4.7}$ per cent of the present-day CMB energy density remains consistent with current observations.

\item The posterior shape in Run~1 is itself informative: rather than resolving to an
interior peak, it rises monotonically and piles up against the prior wall at
$\varepsilon_{\rm CMB} = 1$, with a best-fit value $\varepsilon_{\rm CMB} = {0.9992}$
lying extremely close to unity (Figs.~\ref{fig:corner_run1} and~\ref{fig:posterior_run1}). Making $\varepsilon_{\rm CMB}$ a free parameter to explore the
super-blackbody regime in Run~2
resolves this ambiguity: the posterior is well described
by a Gaussian centred at $\varepsilon_{\rm CMB} = {1.0104^{+0.025}_{-0.024}}$
(Figs.~\ref{fig:corner_run2} and~\ref{fig:posterior_run2}), confirming that the data's
preference for $\varepsilon_{\rm CMB} \approx 1$ is a genuine feature of the posterior
rather than an artefact of the prior boundary. The standard cosmological parameters
recovered in Run~2 (Table~\ref{tab:params_summary}) remain fully consistent with both
the $\Lambda$CDM baseline and Run~1 values.

\item \citetalias{Gjergo2025} estimate a contribution of at least 1.4 per cent from
the formation of massive early-type galaxies at $z \sim 15$-$20$. This falls comfortably
within our 68 per cent credible interval, $\varepsilon_{\rm CMB} \in {[0.971, 0.996]}$,
so the Planck 2018 data are fully consistent with the scenario.

\item The most optimistic \citetalias{Gjergo2025} estimates, in which the grey-body
contribution approaches the full CMB energy density ($\varepsilon_{\rm CMB} \to 0$) , are ruled out at very high significance ($42\sigma$)
by the Run~2 constraint $\varepsilon_{\rm CMB} = {1.0104^{+0.025}_{-0.024}}$.

\item The constraint on $\varepsilon_{\rm CMB}$ does not depend on the assumed transition
redshift. A companion run with $z_t$ left free returns a flat posterior across
$5 \lesssim z_t \lesssim 50$ and an emissivity constraint essentially identical to the
fixed-$z_t$ case: the amplitude of non-primordial contribution is constrained by the CMB
anisotropy spectrum, but its epoch is not.

\item Combined with the independent monopole constraint of
\cite{Corredoira2026_resnote} ($<1.3$ per cent at 95 per cent confidence
for realistic dust parameters), the \citetalias{Gjergo2025} scenario is now bracketed
from two directions: the anisotropy spectrum permits up to
$\sim$5 per cent but cannot probe below $\sim$1.5 per cent, while the
monopole already presses marginally on the conservative 1.4 per cent
prediction. The two probes are complementary rather than redundant:
in the spectrally degenerate limit ($\beta = 0$, $\sigma_d = 0$),
where the monopole is compatible with a CMB of entirely dust origin,
the anisotropy constraint alone excludes that possibility at
$\sim 42\sigma$.

\item None of the standard cosmological parameters shift significantly under the  $\varepsilon_{\rm CMB}$ modification, $H_0$ and $\sigma_8$ in particular are unaffected. Across all
three runs (Table~\ref{tab:params_summary}), the only shift exceeding $1\sigma$,
expressed in units of that run's own posterior uncertainty, is the baryon density
$\omega_b h^2$ in Run~1, at $\sim 1.1\sigma$ relative to the baseline; every other parameter
in each run remains below $1\sigma$. A dust contribution at the \citetalias{Gjergo2025}
level therefore neither helps nor worsens the existing cosmological tensions.

\end{enumerate}

The question addressed here is not unique to the scenario of contamination of CMB photons by thermalization of dust: a long line of non-standard cosmologies, reviewed by \citep{LopezCorredoira2013NonStandard} and by \citep{Cirkovic2017AltCMB} have proposed non-primordial origins for the CMB, and the $\varepsilon_{\rm CMB}$-$z_t$ framework developed in this paper could be applied directly to test them against the same Planck~2018 data set. Because the framework is defined purely in terms of an energy-density contribution above a transition redshift $z_t$, without reference to any specific emission mechanism, it is agnostic to the physical origin of the non-standard component: the same two-parameter test could equally be applied to the achromatic grey-body signature predicted by distance-duality violation (Section~\ref{sec:limitations}) or to other proposed departures from a purely primordial blackbody, using the identical Planck~2018 data set employed here. This is, as far as we are aware, the first observational test of the \citetalias{Gjergo2025} dust emission scenario against the CMB anisotropy power spectrum. The modified \textsc{camb} implementation used here is available on request. Telling standard $\Lambda$CDM apart from a model with a $\sim 1$-$2$ per cent CMB energy density originating from dust emission will need either a real improvement in CMB sensitivity, of the kind expected from the Simons  Observatory \citep{Ade2019SimonsObservatory}, CMB-S4 \citep{Abazajian2016CMBS4}, or LiteBIRD \citep{LiteBIRD2023}, or direct observations of the ETG formation epoch with JWST and future facilities. 

Given how comfortably the current data accommodate the conservative \citetalias{Gjergo2025} prediction, and how decisively they exclude its optimistic extreme, the dust emission acting as a CMB foreground from early-type galaxy formation is best regarded not as a fringe alternative but as a quantitatively bounded, testable component of the microwave sky, whose detection or exclusion is within reach of the next generation of CMB experiments.

\section*{Acknowledgements}

We thank Tirthankar Roy Choudhury (NCRA-TIFR, Pune) and Aseem Paranjape 
(IUCAA, Pune) for valuable discussions and detailed comments on the manuscript. We thank Santanu Das and Lipsa Panda for the useful discussion. We acknowledge the use of the IUCAA HPC cluster Pegasus for the computational work presented in this study. We used artificial intelligence (AI) language assistant tools solely for language editing and improving the clarity of the manuscript. 
\bibliographystyle{JCAP}
\bibliography{akshith_ref}

@article{PenzasWilson1965,
  author  = {Penzias, A.~A. and Wilson, R.~W.},
  title   = {A Measurement of Excess Antenna Temperature at 4080 {Mc/s}},
  journal = {ApJ},
  year    = {1965},
  volume  = {142},
  pages   = {419--421},
  doi     = {10.1086/148307}
}

@article{Fixsen1996,
  author  = {Fixsen, D.~J. and Cheng, E.~S. and Gales, J.~M. and Mather, J.~C.
             and Shafer, R.~A. and Wright, E.~L.},
  title   = {The Cosmic Microwave Background Spectrum from the Full {COBE FIRAS}
             Data Set},
  journal = {ApJ},
  year    = {1996},
  volume  = {473},
  pages   = {576--587},
  doi     = {10.1086/178173}
}

@article{Fixsen2009,
  author  = {Fixsen, D.~J.},
  title   = {The Temperature of the Cosmic Microwave Background},
  journal = {ApJ},
  year    = {2009},
  volume  = {707},
  pages   = {916--920},
  doi     = {10.1088/0004-637X/707/2/916}
}

@article{Bennett2013,
  author  = {Bennett, C.~L. and others},
  title   = {Nine-year Wilkinson Microwave Anisotropy Probe (WMAP) Observations: Final Maps and Results},
  journal = {ApJS},
  year    = {2013},
  volume  = {208},
  pages   = {20},
  doi     = {10.1088/0067-0049/208/2/20}
}

@article{PlanckCollaboration2020,
  author  = {{Planck Collaboration}},
  title   = {{Planck} 2018 results. {VI}. Cosmological parameters},
  journal = {A\&A},
  year    = {2020},
  volume  = {641},
  pages   = {A6},
  doi     = {10.1051/0004-6361/201833910}
}

@article{PlanckLikelihood2020,
  author  = {{Planck Collaboration}},
  title   = {{Planck} 2018 results. {V}. {CMB} power spectra and likelihoods},
  journal = {A\&A},
  year    = {2020},
  volume  = {641},
  pages   = {A5},
  doi     = {10.1051/0004-6361/201936386}
}

@article{PlanckLensing2020,
  author  = {{Planck Collaboration}},
  title   = {{Planck} 2018 results. {VIII}. Gravitational lensing},
  journal = {A\&A},
  year    = {2020},
  volume  = {641},
  pages   = {A8},
  doi     = {10.1051/0004-6361/201833886}
}

@article{Planck2018components,
  author  = {{Planck Collaboration}},
  title   = {{Planck} 2018 results. {IV}. Diffuse component separation},
  journal = {A\&A},
  year    = {2020},
  volume  = {641},
  pages   = {A4},
  doi     = {10.1051/0004-6361/201833881}
}

@article{Peebles1968,
  author  = {Peebles, P.~J.~E. and Yu, J.~T.},
  title   = {Primeval Adiabatic Perturbation in an Expanding Universe},
  journal = {ApJ},
  year    = {1970},
  volume  = {162},
  pages   = {815},
  doi     = {10.1086/150713}
}

@article{Sunyaev1970,
  author  = {Sunyaev, R.~A. and Zel'dovich, Ya.~B.},
  title   = {Small-Scale Fluctuations of Relic Radiation},
  journal = {Ap\&SS},
  year    = {1970},
  volume  = {7},
  pages   = {3--19},
  doi     = {10.1007/BF00653471}
}

@article{Gjergo2025,
  author  = {Gjergo, E. and Kroupa, P.},
  title   = {The Impact of Early Massive Galaxy Formation on the Cosmic Microwave Background},
  journal = {Nuclear Physics B},
  year    = {2025},
  volume  = {1017},
  pages   = {116931},
  doi     = {10.1016/j.nuclphysb.2025.116931}
}

@article{Lewis2000,
  author  = {Lewis, A. and Challinor, A. and Lasenby, A.},
  title   = {Efficient Computation of Cosmic Microwave Background
             Anisotropies in Closed {Friedmann-Robertson-Walker} Models},
  journal = {ApJ},
  year    = {2000},
  volume  = {538},
  pages   = {473--476},
  doi     = {10.1086/309179}
}

@article{Howlett2012,
  author  = {Howlett, C. and Lewis, A. and Hall, A. and Challinor, A.},
  title   = {{CMB} power spectrum parameter degeneracies in the era of
             precision cosmology},
  journal = {JCAP},
  year    = {2012},
  volume  = {04},
  pages   = {027},
  doi     = {10.1088/1475-7516/2012/04/027}
}

@article{Torrado2021,
  author  = {Torrado, J. and Lewis, A.},
  title   = {Cobaya: Code for {Bayesian} Analysis in Cosmology},
  journal = {JCAP},
  year    = {2021},
  volume  = {05},
  pages   = {057},
  doi     = {10.1088/1475-7516/2021/05/057}
}

@article{Gelman1992,
  author  = {Gelman, A. and Rubin, D.~B.},
  title   = {Inference from Iterative Simulation Using Multiple Sequences},
  journal = {Statistical Science},
  year    = {1992},
  volume  = {7},
  pages   = {457--472},
  doi     = {10.1214/ss/1177011136}
}

@article{Beutler2011,
  author  = {Beutler, F. and others},
  title   = {The 6d{F} Galaxy Survey: baryon acoustic oscillations and
             the local {Hubble} constant},
  journal = {MNRAS},
  year    = {2011},
  volume  = {416},
  pages   = {3017--3032},
  doi     = {10.1111/j.1365-2966.2011.19250.x}
}

@article{Ross2015,
  author  = {Ross, A.~J. and others},
  title   = {The clustering of the {SDSS DR7} main Galaxy sample},
  journal = {MNRAS},
  year    = {2015},
  volume  = {449},
  pages   = {835--847},
  doi     = {10.1093/mnras/stv154}
}

@article{Alam2017,
  author  = {Alam, S. and others},
  title   = {The clustering of galaxies in the completed {SDSS-III} Baryon
             Oscillation Spectroscopic Survey: cosmological analysis of the
             {DR12} galaxy sample},
  journal = {MNRAS},
  year    = {2017},
  volume  = {470},
  pages   = {2617--2652},
  doi     = {10.1093/mnras/stx721}
}

@article{Haslbauer2022,
  author  = {Haslbauer, M. and Banik, I. and Kroupa, P. and
             Wittenburg, N. and Javanmardi, B.},
  title   = {The High Fraction of Thin Disk Galaxies Continues to
             Challenge {$\Lambda$CDM} Cosmology},
  journal = {ApJ},
  year    = {2022},
  volume  = {925},
  pages   = {183},
  doi     = {10.3847/1538-4357/ac46ac}
}

@article{Labbe2023,
  author  = {Labb{\'e}, I. and van Dokkum, P. and Nelson, E. and others},
  title   = {A population of red candidate massive galaxies ${\sim}600\,\rm Myr$
             after the {Big Bang}},
  journal = {Nature},
  year    = {2023},
  volume  = {616},
  pages   = {266--269},
  doi     = {10.1038/s41586-023-05786-2}
}

@article{Xiao2024,
  author  = {Xiao, M. and Oesch, P.~A. and Elbaz, D. and others},
  title   = {Accelerated formation of ultra-massive galaxies in the
             first billion years},
  journal = {Nature},
  year    = {2024},
  volume  = {635},
  pages   = {311--315},
  doi     = {10.1038/s41586-024-08094-5}
}

@article{LopezCorredoira2026,
  author  = {L{\'o}pez-Corredoira, M. and Guti{\'e}rrez, C.~M.},
  title   = {Improved measurements of the age of {JWST} galaxies at
             $z = 6$--$10$},
  journal = {MNRAS},
  year    = {2026},
  volume  = {546},
  number  = {2},
  eid     = {stag089},
  pages   = {stag089},
  doi     = {10.1093/mnras/stag089}
}

@article{Kogut2011,
  author  = {Kogut, A. and others},
  title   = {The Primordial Inflation Explorer ({PIXIE}): A Nulling
             Polarimeter for Cosmic Microwave Background Observations},
  journal = {JCAP},
  year    = {2011},
  volume  = {07},
  pages   = {025},
  doi     = {10.1088/1475-7516/2011/07/025}
}

@article{Mather1994,
  author  = {Mather, J.~C. and Cheng, E.~S. and Cottingham, D.~A.
             and Eplee, R.~E. and Fixsen, D.~J. and Hewagama, T.
             and Isaacman, R.~B. and Jensen, K.~A. and Meyer, S.~S.
             and Noerdlinger, P.~D. and Read, S.~M. and Rosen, L.~P.
             and Shafer, R.~A. and Wright, E.~L. and Bennett, C.~L.
             and Boggess, N.~W. and Hauser, M.~G. and Kelsall, T.
             and Moseley, S.~H. and Silverberg, R.~F. and Smoot, G.~F.
             and Weiss, R. and Wilkinson, D.~T.},
  title   = {Measurement of the Cosmic Microwave Background Spectrum
             by the {COBE FIRAS} Instrument},
  journal = {ApJ},
  year    = {1994},
  volume  = {420},
  pages   = {439--444},
  doi     = {10.1086/173574}
}

@article{Puget1996,
  author  = {Puget, J.-L. and Abergel, A. and Bernard, J.-P.
             and Boulanger, F. and Burton, W.~B. and D{\'e}sert, F.-X.
             and Hartmann, D.},
  title   = {Tentative detection of a cosmic far-infrared background
             with {COBE}},
  journal = {A\&A},
  year    = {1996},
  volume  = {308},
  pages   = {L5--L8}
}

@article{Fixsen1998,
  author  = {Fixsen, D.~J. and Dwek, E. and Mather, J.~C.
             and Bennett, C.~L. and Shafer, R.~A.},
  title   = {The Spectrum of the Extragalactic Far-Infrared Background
             from the {COBE FIRAS} Observations},
  journal = {ApJ},
  year    = {1998},
  volume  = {508},
  pages   = {123--128},
  doi     = {10.1086/306383}
}

@article{HauserDwek2001,
  author  = {Hauser, M.~G. and Dwek, E.},
  title   = {The Cosmic Infrared Background: Measurements and Implications},
  journal = {ARA\&A},
  year    = {2001},
  volume  = {39},
  pages   = {249--307},
  doi     = {10.1146/annurev.astro.39.1.249}
}

@article{KS19,
  author  = {Khaire, V. and Srianand, R.},
  title   = {New synthesis models of consistent extragalactic background
             light over cosmic time},
  journal = {MNRAS},
  year    = {2019},
  volume  = {484},
  pages   = {4174--4199},
  doi     = {10.1093/mnras/stz174}
}

@ARTICLE{Srianandmain,
       author = {{Srianand}, R. and {Petitjean}, P. and {Ledoux}, C.},
        title = "{The cosmic microwave background radiation temperature at a redshift of 2.34}",
      journal = {Nature},
         year = 2000,
       volume = {408},
       number = {6815},
        pages = {931-935},
          doi = {10.1038/35050020},
archivePrefix = {arXiv},
       eprint = {astro-ph/0012222},
 primaryClass = {astro-ph},
       adsurl = {https://ui.adsabs.harvard.edu/abs/2000Natur.408..931S}
}

@ARTICLE{Noterdaeme-Srianand,
       author = {{Noterdaeme}, P. and {Petitjean}, P. and {Srianand}, R. and {Ledoux}, C. and {L{\'o}pez}, S.},
        title = "{The evolution of the cosmic microwave background temperature. Measurements of T$_{CMB}$ at high redshift from carbon monoxide excitation}",
      journal = {Astronomy \& Astrophysics},
         year = 2011,
       volume = {526},
          eid = {L7},
        pages = {L7},
          doi = {10.1051/0004-6361/201016140},
archivePrefix = {arXiv},
       eprint = {1012.3164},
 primaryClass = {astro-ph.CO},
       adsurl = {https://ui.adsabs.harvard.edu/abs/2011A&A...526L...7N}
}

@article{Khaire15ebl,
  author        = {Khaire, Vikram and Srianand, Raghunathan},
  title         = {Star Formation History, Dust Attenuation, and Extragalactic Background Light},
  journal       = {ApJ},
  year          = {2015},
  volume        = {805},
  number        = {1},
  pages         = {33},
  doi           = {10.1088/0004-637X/805/1/33},
  eprint        = {1405.7038},
  archivePrefix = {arXiv},
  primaryClass  = {astro-ph.GA},
}

@article{BoylanKolchin2023,
  author  = {Boylan-Kolchin, M.},
  title   = {Stress testing {$\Lambda$CDM} with high-redshift galaxy
             candidates},
  journal = {Nature Astronomy},
  year    = {2023},
  volume  = {7},
  pages   = {731--735},
  doi     = {10.1038/s41550-023-01937-7}
}

@article{Riess2022,
  author  = {Riess, A.~G. and others},
  title   = {A Comprehensive Measurement of the Local Value of the
             {Hubble} Constant with 1 km/s/Mpc Uncertainty from the
             {Hubble Space Telescope} and the {SH0ES} Team},
  journal = {ApJL},
  year    = {2022},
  volume  = {934},
  pages   = {L7},
  doi     = {10.3847/2041-8213/ac5c5b}
}

@article{Verde2019,
  author  = {Verde, L. and Treu, T. and Riess, A.~G.},
  title   = {Tensions between the Early and the Late Universe},
  journal = {Nature Astronomy},
  year    = {2019},
  volume  = {3},
  pages   = {891--895},
  doi     = {10.1038/s41550-019-0902-0}
}

@article{DiValentino2021,
  author  = {Di Valentino, E. and others},
  title   = {In the Realm of the {Hubble} Tension --- a Review of Solutions},
  journal = {Classical and Quantum Gravity},
  year    = {2021},
  volume  = {38},
  pages   = {153001},
  doi     = {10.1088/1361-6382/ac086d}
}

@article{Chluba2021SpectralDistortions,
  author  = {Chluba, J. and Abitbol, M.~H. and Aghanim, N. and Ali-Ha{\"i}moud, Y. and Alvarez, M. and Basu, K. and Bolliet, B. and Burigana, C. and de Bernardis, P. and Delabrouille, J. and Dimastrogiovanni, E. and Finelli, F. and Fixsen, D. and Hart, L. and Hern{\'a}ndez-Monteagudo, C. and Hill, J.~C. and Kogut, A. and Kohri, K. and Lesgourgues, J. and Maffei, B. and Mather, J. and Mukherjee, S. and Patil, S.~P. and Ravenni, A. and Remazeilles, M. and Rotti, A. and Rubi{\~n}o-Mart{\'i}n, J.~A. and Silk, J. and Sunyaev, R.~A. and Switzer, E.~R.},
  title   = {{New Horizons in Cosmology with Spectral Distortions of the Cosmic Microwave Background}},
  journal = {Experimental Astronomy},
  year    = {2021},
  volume  = {51},
  number  = {3},
  pages   = {1515--1554},
  note    = {arXiv:1909.01593},
  doi     = {10.1007/s10686-021-09729-5}
}

@article{ChlubaSunyaev2012,
  author  = {Chluba, J. and Sunyaev, R.~A.},
  title   = {{The evolution of CMB spectral distortions in the early Universe}},
  journal = {MNRAS},
  year    = {2012},
  volume  = {419},
  pages   = {1294--1314},
  note    = {arXiv:1109.6552},
  doi     = {10.1111/j.1365-2966.2011.19786.x}
}

@article{ChlubaKhatriSunyaev2012,
  author  = {Chluba, J. and Khatri, R. and Sunyaev, R.~A.},
  title   = {{CMB at 2x2 order: the dissipation of primordial acoustic waves and the observable part of the associated energy release}},
  journal = {MNRAS},
  year    = {2012},
  volume  = {425},
  pages   = {1129--1169},
  note    = {arXiv:1202.0057},
  doi     = {10.1111/j.1365-2966.2012.21474.x}
}

@article{KhatriSunyaev2012,
  author  = {Khatri, R. and Sunyaev, R.~A.},
  title   = {{Creation of the CMB spectrum: precise analytic solutions for the blackbody photosphere}},
  journal = {JCAP},
  year    = {2012},
  volume  = {06},
  pages   = {038},
  note    = {arXiv:1203.2601},
  doi     = {10.1088/1475-7516/2012/06/038}
}

@article{Chluba2016,
  author  = {Chluba, J.},
  title   = {{Which spectral distortions does {$\Lambda$}CDM actually predict?}},
  journal = {MNRAS},
  year    = {2016},
  volume  = {460},
  pages   = {227--239},
  note    = {arXiv:1603.02496},
  doi     = {10.1093/mnras/stw945}
}

@article{Chluba2019Astro2020,
  author  = {Chluba, J. and Kogut, A. and Patil, S.~P. and Abitbol, M.~H. and Aghanim, N. and Ali-Ha{\"i}moud, Y. and Amin, M.~A. and others},
  title   = {{Spectral Distortions of the CMB as a Probe of Inflation, Recombination, Structure Formation and Particle Physics: Astro2020 Science White Paper}},
  journal = {Bulletin of the AAS},
  year    = {2019},
  volume  = {51},
  pages   = {184},
  note    = {arXiv:1903.04218}
}

@article{Kogut2025PIXIE,
  author  = {Kogut, A. and Aghanim, N. and Chluba, J. and Chuss, D.~T. and Delabrouille, J. and Dvorkin, C. and Fixsen, D. and Ghosh, S. and Hensley, B.~S. and Hill, J.~C. and Maffei, B. and Pullen, A.~R. and Rotti, A. and Sabyr, A. and Switzer, E.~R. and Thiele, L. and Wollack, E.~J. and Zelko, I.},
  title   = {{The Primordial Inflation Explorer (PIXIE): Mission Design and Science Goals}},
  journal = {JCAP},
  year    = {2025},
  volume  = {04},
  pages   = {020},
  note    = {arXiv:2405.20403},
  doi     = {10.1088/1475-7516/2025/04/020}
}

@article{Louis2025ACTDR6,
  author  = {Louis, T. and others},
  title   = {{The Atacama Cosmology Telescope: DR6 Power Spectra, Likelihoods and {$\Lambda$}CDM Parameters}},
  journal = {JCAP},
  year    = {2025},
  volume  = {11},
  pages   = {062},
  note    = {arXiv:2503.14452},
  doi     = {10.1088/1475-7516/2025/11/062}
}

@article{Calabrese2025ACTDR6Extended,
  author  = {Calabrese, E. and others},
  title   = {{The Atacama Cosmology Telescope: DR6 Constraints on Extended Cosmological Models}},
  journal = {JCAP},
  year    = {2025},
  volume  = {11},
  pages   = {063},
  doi     = {10.1088/1475-7516/2025/11/063},
  note    = {arXiv:2503.14454}
}

@article{Madhavacheril2024ACTDR6Lensing,
  author  = {Madhavacheril, M.~S. and Qu, F.~J. and Sherwin, B.~D. and MacCrann, N. and Li, Y. and Abril-Cabezas, I. and others},
  title   = {{The Atacama Cosmology Telescope: DR6 Gravitational Lensing Map and Cosmological Parameters}},
  journal = {ApJ},
  year    = {2024},
  volume  = {962},
  number  = {2},
  pages   = {113},
  note    = {arXiv:2304.05203},
  doi     = {10.3847/1538-4357/acff5f}
}

@article{Qu2024ACTDR6LensingPowerSpectrum,
  author  = {Qu, F.~J. and Sherwin, B.~D. and Madhavacheril, M.~S. and others},
  title   = {{The Atacama Cosmology Telescope: A Measurement of the DR6 CMB Lensing Power Spectrum and Its Implications for Structure Growth}},
  journal = {ApJ},
  year    = {2024},
  volume  = {962},
  number  = {2},
  pages   = {112},
  note    = {arXiv:2304.05202}
}

@article{Beringue2025ACTDR6Foregrounds,
  author  = {Beringue, B. and Surrao, K.~M. and Hill, J.~C. and Atkins, Z. and Battaglia, N. and Bolliet, B. and Calabrese, E. and Choi, S.~K. and Clark, S.~E. and Duivenvoorden, A.~J. and Dunkley, J. and Giardiello, S. and Goldstein, S. and Hensley, B.~S. and Hlo{\v{z}}ek, R. and Jense, H.~T. and Kramer, D. and La Posta, A. and Louis, T. and Mehta, Y. and Moodley, K. and Naess, S. and Partridge, B. and Qu, F.~J. and Ried Guachalla, B. and Sehgal, N. and Sif{\'o}n, C. and Staggs, S.~T. and Trac, H. and Van Engelen, A. and Wollack, E.~J.},
  title   = {{The Atacama Cosmology Telescope: DR6 Power Spectrum Foreground Model and Validation}},
  journal = {JCAP},
  year    = {2025},
  volume  = {10},
  pages   = {082},
  note    = {arXiv:2506.06274},
  doi     = {10.1088/1475-7516/2025/10/082}
}

@article{AbdulKarim2025DESIDR2,
  author  = {{Abdul Karim}, M. and others},
  title   = {{DESI DR2 Results II: Measurements of Baryon Acoustic Oscillations and Cosmological Constraints}},
  journal = {Phys. Rev. D},
  year    = {2025},
  volume  = {112},
  pages   = {083515},
  note    = {arXiv:2503.14738},
  doi     = {10.1103/PhysRevD.112.083515}
}

@article{Shen2024EarlyDarkEnergy,
  author  = {Shen, X. and Vogelsberger, M. and Boylan-Kolchin, M. and Tacchella, S. and Naidu, R.~P.},
  title   = {{Early Galaxies and Early Dark Energy: A Unified Solution to the Hubble Tension and Puzzles of Massive Bright Galaxies revealed by JWST}},
  journal = {MNRAS},
  year    = {2024},
  volume  = {533},
  number  = {4},
  pages   = {3923--3936},
  note    = {arXiv:2406.15548},
  doi     = {10.1093/mnras/stae1932}
}

@article{KrishnanAbazajian2026,
  author  = {Krishnan, J.~R. and Abazajian, K.~N.},
  title   = {{Statistics Meet Systematics: Resolution of the Massive Early JWST Galaxy Tension}},
  journal = {Phys. Rev. D},
  year    = {2026},
  volume  = {113},
  pages   = {083007},
  note    = {arXiv:2511.13708},
  doi     = {10.1103/PhysRevD.113.083007}
}

@article{Cyburt2016BBN,
  author  = {Cyburt, R.~H. and Fields, B.~D. and Olive, K.~A. and Yeh, T.-H.},
  title   = {{Big Bang Nucleosynthesis: 2015}},
  journal = {Rev. Mod. Phys.},
  year    = {2016},
  volume  = {88},
  pages   = {015004},
  note    = {arXiv:1505.01076},
  doi     = {10.1103/RevModPhys.88.015004}
}

@article{CaseyNarayananCooray2014,
  author  = {Casey, C.~M. and Narayanan, D. and Cooray, A.},
  title   = {{Dusty Star-Forming Galaxies at High Redshift}},
  journal = {Physics Reports},
  year    = {2014},
  volume  = {541},
  number  = {2},
  pages   = {45--161},
  note    = {arXiv:1402.1456},
  doi     = {10.1016/j.physrep.2014.02.009}
}

@article{Bethermin2012,
  author  = {B{\'e}thermin, M. and Daddi, E. and Magdis, G. and Sargent, M.~T. and Hezaveh, Y. and Elbaz, D. and Le Borgne, D. and Mullaney, J. and Pannella, M. and Buat, V. and Charmandaris, V. and Lagache, G. and Scott, D.},
  title   = {{A Unified Empirical Model for Infrared Galaxy Counts Based on the Observed Physical Evolution of Distant Galaxies}},
  journal = {ApJL},
  year    = {2012},
  volume  = {757},
  number  = {2},
  pages   = {L23},
  note    = {arXiv:1208.6512},
  doi     = {10.1088/2041-8205/757/2/L23}
}

@article{AguirreHaiman2000,
  author  = {Aguirre, A. and Haiman, Z.},
  title   = {Cosmological Constant or Intergalactic Dust? Constraints from the Cosmic Far-Infrared Background},
  journal = {ApJ}, volume = {532}, pages = {28--36}, year = {2000},
  doi     = {10.1086/308557}}

@article{Cirkovic2017AltCMB,
  author  = {\'Cirkovi\'c, M. M. and Perovi\'c, S.},
  title   = {Alternative Explanations of the Cosmic Microwave Background: A Historical and an Epistemological Perspective},
  journal = {Stud. Hist. Philos. Mod. Phys.}, volume = {62}, pages = {1--18}, year = {2018},
  doi     = {10.1016/j.shpsb.2017.04.005},
  note    = {arXiv:1705.07721}}

@article{Gupta2023CCCTL,
  author  = {Gupta, R. P.},
  title   = {JWST early Universe observations and $\Lambda$CDM cosmology},
  journal = {MNRAS}, volume = {524}, number = {3}, pages = {3385--3395}, year = {2023},
  doi     = {10.1093/mnras/stad2032}}

@misc{Moffat2024MOG,
  author = {Moffat, J. W.},
  title  = {Galaxy Formation in the Early Universe},
  year   = {2024}, note = {arXiv:2412.03534}}

@misc{vanPutten2023FastFurious,
  author = {van Putten, M. H. P. M.},
  title  = {The Fast and Furious in JWST high-$z$ galaxies},
  year   = {2023}, note = {arXiv:2312.16692}}

@article{LopezCorredoira2013NonStandard,
  author  = {L{\'o}pez-Corredoira, M.},
  title   = {Non-standard models and the sociology of cosmology},
  journal = {Stud. Hist. Philos. Mod. Phys.}, volume = {46}, pages = {86--96}, year = {2014},
  note    = {arXiv:1305.5844}}

@misc{Fabbian2025,
  author = {Fabbian, Giulio and Bianchini, Federico and Sabyr, Alina and
            Hill, J. Colin and Lovell, Christopher C. and
            Thiele, Leander and Spergel, David N.},
  title = {A new constraint on the $y$-distortion with FIRAS: implications for feedback models in galaxy formation and cosmic shear measurements},
  year = {2025},
  eprint = {2512.03038},
  archivePrefix = {arXiv},
  primaryClass = {astro-ph.CO},
  note = {arXiv:2512.03038}
}

@article{Ellis2013,
  author  = {Ellis, George F. R. and Poltis, Robert and Uzan, Jean-Philippe and Weltman, Amanda},
  title   = {The blackness of the cosmic microwave background spectrum as a probe of the distance-duality relation},
  journal = {Phys. Rev. D},
  volume  = {87},
  pages   = {103530},
  year    = {2013},
  eprint  = {1301.1312},
  archivePrefix = {arXiv},
  primaryClass  = {astro-ph.CO},
  doi     = {10.1103/PhysRevD.87.103530}
}

@article{Eggen1962,
  author  = {Eggen, O. J. and Lynden-Bell, D. and Sandage, A. R.},
  title   = {Evidence from the motions of old stars that the Galaxy
             collapsed},
  journal = {ApJ},
  year    = {1962},
  volume  = {136},
  pages   = {748--766},
  doi     = {10.1086/147433}
}

@article{Larson1974,
  author  = {Larson, R. B.},
  title   = {Dynamical models for the formation and evolution of
             elliptical galaxies},
  journal = {MNRAS},
  year    = {1974},
  volume  = {166},
  pages   = {585--616},
  doi     = {10.1093/mnras/166.3.585}
}

@article{Seager2000,
  author  = {Seager, S. and Sasselov, D. D. and Scott, D.},
  title   = {How Exactly Did the Universe Become Neutral?},
  journal = {ApJS},
  year    = {2000},
  volume  = {128},
  pages   = {407--430},
  doi     = {10.1086/313388}
}

@misc{Abazajian2016CMBS4,
  author  = {Abazajian, K. N. and others},
  title   = {CMB-S4 Science Book, First Edition},
  year    = {2016},
  eprint  = {1610.02743},
  archivePrefix = {arXiv},
  primaryClass  = {astro-ph.CO},
  note    = {arXiv:1610.02743}
}

@article{Ade2019SimonsObservatory,
  author  = {Ade, P. and others},
  title   = {The Simons Observatory: Science Goals and Forecasts},
  journal = {JCAP},
  volume  = {2019},
  pages   = {056},
  year    = {2019},
  doi     = {10.1088/1475-7516/2019/02/056},
  note    = {arXiv:1808.07445}
}

@ARTICLE{SPT3G2025,
       author = {{Camphuis}, E. and {Quan}, W. and {Balkenhol}, L. and {Khalife}, A.~R. and {Ge}, F. and {Guidi}, F. and {Huang}, N. and {Lynch}, G.~P. and {Omori}, Y. and {Trendafilova}, C. and {Anderson}, A.~J. and {Ansarinejad}, B. and {Archipley}, M. and {Barry}, P.~S. and {Benabed}, K. and {Bender}, A.~N. and {Benson}, B.~A. and {Bianchini}, F. and {Bleem}, L.~E. and {Bouchet}, F.~R. and {Bryant}, L. and {Campitiello}, M.~G. and {Carlstrom}, J.~E. and {Chang}, C.~L. and {Chaubal}, P. and {Chichura}, P.~M. and {Chokshi}, A. and {Chou}, T.-L. and {Coerver}, A. and {Crawford}, T.~M. and {Daley}, C. and {de Haan}, T. and {Dibert}, K.~R. and {Dobbs}, M.~A. and {Doohan}, M. and {Doussot}, A. and {Dutcher}, D. and {Everett}, W. and {Feng}, C. and {Ferguson}, K.~R. and {Fichman}, K. and {Foster}, A. and {Galli}, S. and {Gambrel}, A.~E. and {Gardner}, R.~W. and {Goeckner-Wald}, N. and {Gualtieri}, R. and {Guns}, S. and {Halverson}, N.~W. and {Hivon}, E. and {Holder}, G.~P. and {Holzapfel}, W.~L. and {Hood}, J.~C. and {Hryciuk}, A. and {K{\'e}ruzor{\'e}}, F. and {Knox}, L. and {Korman}, M. and {Kornoelje}, K. and {Kuo}, C.-L. and {Levy}, K. and {Lowitz}, A.~E. and {Lu}, C. and {Maniyar}, A. and {Martsen}, E.~S. and {Menanteau}, F. and {Millea}, M. and {Montgomery}, J. and {Nakato}, Y. and {Natoli}, T. and {Noble}, G.~I. and {Ouellette}, A. and {Pan}, Z. and {Paschos}, P. and {Phadke}, K.~A. and {Pollak}, A.~W. and {Prabhu}, K. and {Raghunathan}, S. and {Rahimi}, M. and {Rahlin}, A. and {Reichardt}, C.~L. and {Rouble}, M. and {Ruhl}, J.~E. and {Schiappucci}, E. and {Simpson}, A. and {Sobrin}, J.~A. and {Stark}, A.~A. and {Stephen}, J. and {Tandoi}, C. and {Thorne}, B. and {Umilta}, C. and {Vieira}, J.~D. and {Vitrier}, A. and {Wan}, Y. and {Whitehorn}, N. and {Wu}, W.~L.~K. and {Young}, M.~R. and {Zebrowski}, J.~A. and {SPT-3G Collaboration}},
        title = "{SPT-3G D1: CMB temperature and polarization power spectra and cosmology from 2019 and 2020 observations of the SPT-3G main field}",
      journal = {Physical Review D},
         year = 2026,
       volume = {113},
       number = {8},
          eid = {083504},
        pages = {083504},
          doi = {10.1103/7wt3-9v2y},
archivePrefix = {arXiv},
       eprint = {2506.20707},
 primaryClass = {astro-ph.CO},
       adsurl = {https://ui.adsabs.harvard.edu/abs/2026PhRvD.113h3504C}
}

@article{Melia2022A,
  author  = {Melia, F.},
  title   = {A Population III--Generated Dust Screen at z $\sim$ 16},
  journal = {The Astrophysical Journal},
  year    = {2022},
  volume  = {941},
  pages   = {33},
  doi     = {10.3847/1538-4357/aca412},
}

@article{Wesson1975The,
title={The interrelationship between cosmic dust and the microwave background},
author={P. Wesson},
journal={Astrophysics and Space Science},
year={1975},
volume={36},
pages={363-382},
doi={10.1007/bf00645261}
}

@article{Wickramasinghe1975A,
title={A dust model for the cosmic microwave background},
author={N. Wickramasinghe and M. Edmunds and S. Chitre and J. Narlikar and S. Ramadurai},
journal={Astrophysics and Space Science},
year={1975},
volume={35},
pages={L9-L13},
doi={10.1007/bf00644839}
}

@article{LiteBIRD2023,
    author = "{LiteBIRD Collaboration} and Allys, E. and others",
    title = "{Probing Cosmic Inflation with the LiteBIRD Cosmic Microwave Background Polarization Survey}",
    journal = "PTEP",
    volume = "2023",
    number = "4",
    pages = "042F01",
    year = "2023",
    eprint = "2202.02773",
    archivePrefix = "arXiv",
    primaryClass = "astro-ph.IM",
    doi = "10.1093/ptep/ptac150"
}

@ARTICLE{Corredoira2026_resnote,
       author = {{L{\'o}pez-Corredoira}, Mart{\'\i}n},
        title = "{Constraints on CMBR Flux due to High-z Dust Emission}",
      journal = {Research Notes of the American Astronomical Society},
         year = 2026,
       volume = {10},
       number = {2},
          eid = {40},
        pages = {40},
          doi = {10.3847/2515-5172/ae497b},
archivePrefix = {arXiv},
       eprint = {2602.20916},
 primaryClass = {astro-ph.CO},
       adsurl = {https://ui.adsabs.harvard.edu/abs/2026RNAAS..10...40L}
}

@article{Wong2020H0LiCOW,
  author  = {Wong, K.~C. and others},
  title   = {{H0LiCOW XIII.} A {2.4}\% measurement of {$H_0$} from lensed
             quasars: {5.3$\sigma$} tension between early- and late-Universe
             probes},
  journal = {MNRAS},
  year    = {2020},
  volume  = {498},
  pages   = {1420--1439},
  doi     = {10.1093/mnras/stz3094},
  eprint  = {1907.04869},
  archivePrefix = {arXiv},
  primaryClass  = {astro-ph.CO}
}

@article{Heymans2021KiDS1000,
  author  = {Heymans, C. and others},
  title   = {{KiDS-1000} Cosmology: Multi-probe weak gravitational lensing
             and spectroscopic galaxy clustering constraints},
  journal = {A\&A},
  year    = {2021},
  volume  = {646},
  pages   = {A140},
  doi     = {10.1051/0004-6361/202039063},
  eprint  = {2007.15632},
  archivePrefix = {arXiv},
  primaryClass  = {astro-ph.CO}
}

@article{DESCollaboration2022Y3,
  author  = {{DES Collaboration}},
  title   = {Dark Energy Survey Year 3 Results: Cosmological Constraints
             from Galaxy Clustering and Weak Lensing},
  journal = {Phys. Rev. D},
  year    = {2022},
  volume  = {105},
  pages   = {023520},
  doi     = {10.1103/PhysRevD.105.023520},
  eprint  = {2105.13549},
  archivePrefix = {arXiv},
  primaryClass  = {astro-ph.CO}
}

@article{DiValentino2021S8,
  author  = {Di Valentino, E. and others},
  title   = {Cosmology Intertwined {III}: {f}$\sigma_8$ and {$S_8$}},
  journal = {Astroparticle Physics},
  year    = {2021},
  volume  = {131},
  pages   = {102604},
  doi     = {10.1016/j.astropartphys.2021.102604},
  eprint  = {2008.11285},
  archivePrefix = {arXiv},
  primaryClass  = {astro-ph.CO}
}

@article{Zwicky1929TiredLight,
  author  = {Zwicky, F.},
  title   = {On the Red Shift of Spectral Lines through Interstellar Space},
  journal = {Proc. Natl. Acad. Sci.},
  year    = {1929},
  volume  = {15},
  pages   = {773--779},
  doi     = {10.1073/pnas.15.10.773}
}

@article{MeliaShevchuk2012RhCT,
  author  = {Melia, F. and Shevchuk, A. S. H.},
  title   = {The {$R_h = ct$} Universe},
  journal = {MNRAS},
  year    = {2012},
  volume  = {419},
  pages   = {2579--2586},
  doi     = {10.1111/j.1365-2966.2011.19906.x},
  eprint  = {1109.5189},
  archivePrefix = {arXiv},
  primaryClass  = {astro-ph.CO}
}

@article{Melia2018RhCTReview,
  author  = {Melia, F.},
  title   = {The {$R_h = ct$} Universe Without Inflation},
  journal = {A\&A},
  year    = {2013},
  volume  = {553},
  pages   = {A76},
  doi     = {10.1051/0004-6361/201220447},
  eprint  = {1304.4599},
  archivePrefix = {arXiv},
  primaryClass  = {astro-ph.CO}
}

@article{HoyleBurbidgeNarlikar1993QSSC,
  author  = {Hoyle, F. and Burbidge, G. and Narlikar, J. V.},
  title   = {A Quasi-Steady State Cosmological Model with Creation of Matter},
  journal = {ApJ},
  year    = {1993},
  volume  = {410},
  pages   = {437--457},
  doi     = {10.1086/172761}
}

@article{SachsWolfe1967,
  author  = {Sachs, R.~K. and Wolfe, A.~M.},
  title   = {Perturbations of a Cosmological Model and Angular Variations
             of the Microwave Background},
  journal = {ApJ},
  year    = {1967},
  volume  = {147},
  pages   = {73--90},
  doi     = {10.1086/148982}
}

@article{Harris2020numpy,
  author  = {Harris, C. R. and Millman, K. J. and van der Walt, S. J. and
             Gommers, R. and Virtanen, P. and Cournapeau, D. and
             Wieser, E. and Taylor, J. and Berg, S. and Smith, N. J.
             and Kern, R. and Picus, M. and Hoyer, S. and van Kerkwijk,
             M. H. and Brett, M. and Haldane, A. and del R{\'i}o, J. F.
             and Wiebe, M. and Peterson, P. and G{\'e}rard-Marchant, P.
             and Sheppard, K. and Reddy, T. and Weckesser, W. and
             Abbasi, H. and Gohlke, C. and Oliphant, T. E.},
  title   = {Array Programming with {NumPy}},
  journal = {Nature},
  year    = {2020},
  volume  = {585},
  pages   = {357--362},
  doi     = {10.1038/s41586-020-2649-2}
}

@article{Virtanen2020scipy,
  author  = {Virtanen, P. and Gommers, R. and Oliphant, T. E. and
             Haberland, M. and Reddy, T. and Cournapeau, D. and
             Burovski, E. and Peterson, P. and Weckesser, W. and
             Bright, J. and van der Walt, S. J. and Brett, M. and
             Wilson, J. and Millman, K. J. and Mayorov, N. and
             Nelson, A. R. J. and Jones, E. and Kern, R. and
             Larson, E. and Carey, C. J. and Polat, {\.I}. and
             Feng, Y. and Moore, E. W. and VanderPlas, J. and
             Laxalde, D. and Perktold, J. and Cimrman, R. and
             Henriksen, I. and Quintero, E. A. and Harris, C. R. and
             Archibald, A. M. and Ribeiro, A. H. and Pedregosa, F.
             and van Mulbregt, P. and {SciPy 1.0 Contributors}},
  title   = {{SciPy} 1.0: Fundamental Algorithms for Scientific
             Computing in {Python}},
  journal = {Nature Methods},
  year    = {2020},
  volume  = {17},
  pages   = {261--272},
  doi     = {10.1038/s41592-019-0686-2}
}

@article{ForemanMackey2016corner,
  author  = {Foreman-Mackey, D.},
  title   = {corner.py: Scatterplot matrices in {Python}},
  journal = {Journal of Open Source Software},
  year    = {2016},
  volume  = {1},
  number  = {2},
  pages   = {24},
  doi     = {10.21105/joss.00024}
}

@article{Fan2006,
  author  = {Fan, X. and Strauss, M. A. and Becker, R. H. and White, R. L.
             and Gunn, J. E. and Knapp, G. R. and Richards, G. T. and
             Schneider, D. P. and Brinkmann, J. and Fukugita, M.},
  title   = {Constraining the Evolution of the Ionizing Background and
             the Epoch of Reionization with {z}$\sim$6 Quasars. {II}.
             A Sample of 19 Quasars},
  journal = {AJ},
  year    = {2006},
  volume  = {132},
  pages   = {117--136},
  doi     = {10.1086/504836},
  eprint  = {astro-ph/0512082},
  archivePrefix = {arXiv},
  primaryClass  = {astro-ph}
}

@article{Robertson2015,
  author  = {Robertson, B. E. and Ellis, R. S. and Furlanetto, S. R. and
             Dunlop, J. S.},
  title   = {Cosmic Reionization and Early Star-forming Galaxies: A
             Joint Analysis of new Constraints from {Planck} and the
             {Hubble} Space Telescope},
  journal = {ApJL},
  year    = {2015},
  volume  = {802},
  pages   = {L19},
  doi     = {10.1088/2041-8205/802/2/L19},
  eprint  = {1502.02024},
  archivePrefix = {arXiv},
  primaryClass  = {astro-ph.CO}
}

@misc{Carlos25_CMB_tired_light,
  author        = {Carlos Navia},
  title         = {Cosmic Microwave Background Radiation within the Zwicky Tired Light Hypothesis},
  year          = {2025},
  eprint        = {2504.10510},
  archivePrefix = {arXiv},
  primaryClass  = {physics.gen-ph},
  doi           = {10.48550/arXiv.2504.10510},
}

@ARTICLE{Narlikar2002Inhomogeneities,
       author = {{Narlikar}, J.~V. and {Vishwakarma}, R.~G. and {Hajian}, Amir and {Souradeep}, Tarun and {Burbidge}, G. and {Hoyle}, F.},
        title = "{Inhomogeneities in the Microwave Background Radiation Interpreted within the Framework of the Quasi-Steady State Cosmology}",
      journal = {Astrophysical Journal},
         year = 2003,
       volume = {585},
       number = {1},
        pages = {1-11},
          doi = {10.1086/345928},
archivePrefix = {arXiv},
       eprint = {astro-ph/0211036},
 primaryClass = {astro-ph},
       adsurl = {https://ui.adsabs.harvard.edu/abs/2003ApJ...585....1N}
}

@ARTICLE{Fujii20_CMB_Rh_ct,
       author = {{Fujii}, Hirokazu},
        title = "{Inconsistency of the R$_{h}$ = ct Cosmology from the Viewpoint of the Redshift of the Cosmic Microwave Background Radiation}",
      journal = {Research Notes of the American Astronomical Society},
         year = 2020,
       volume = {4},
       number = {5},
          eid = {72},
        pages = {72},
          doi = {10.3847/2515-5172/ab9537},
       adsurl = {https://ui.adsabs.harvard.edu/abs/2020RNAAS...4...72F}
}

@ARTICLE{Chiara26_distance_duality_DESI,
       author = {{Alfano}, Anna Chiara and {Luongo}, Orlando},
        title = "{Cosmic distance duality after DESI 2024 data release and dark energy evolution}",
      journal = {Physics of the Dark Universe},
         year = 2026,
       volume = {51},
          eid = {102205},
        pages = {102205},
          doi = {10.1016/j.dark.2025.102205},
archivePrefix = {arXiv},
       eprint = {2501.15233},
 primaryClass = {astro-ph.CO},
       adsurl = {https://ui.adsabs.harvard.edu/abs/2026PDU....5102205A}
}

@article{BashinskySeljak2004,
  author  = {Bashinsky, S. and Seljak, U.},
  title   = {Neutrino perturbations in {CMB} anisotropy and matter clustering},
  journal = {Phys. Rev. D},
  year    = {2004},
  volume  = {69},
  pages   = {083002},
  doi     = {10.1103/PhysRevD.69.083002},
  eprint  = {astro-ph/0310198},
  archivePrefix = {arXiv},
  primaryClass  = {astro-ph}
}

@article{Follin2015,
  author  = {Follin, B. and Knox, L. and Millea, M. and Pan, Z.},
  title   = {First Detection of the Acoustic Oscillation Phase Shift Expected
             from the Cosmic Neutrino Background},
  journal = {Phys. Rev. Lett.},
  year    = {2015},
  volume  = {115},
  pages   = {091301},
  doi     = {10.1103/PhysRevLett.115.091301},
  eprint  = {1503.07863},
  archivePrefix = {arXiv},
  primaryClass  = {astro-ph.CO}
}

@article{BaumannGreenWallisch2016,
  author  = {Baumann, D. and Green, D. and Wallisch, B.},
  title   = {New Target for Cosmic Axion Searches},
  journal = {Phys. Rev. Lett.},
  year    = {2016},
  volume  = {117},
  pages   = {171301},
  doi     = {10.1103/PhysRevLett.117.171301},
  eprint  = {1604.08614},
  archivePrefix = {arXiv},
  primaryClass  = {astro-ph.CO}
}

@book{Adams1979,
  author    = {Adams, Douglas},
  title     = {The Hitchhiker's Guide to the Galaxy},
  publisher = {Pan Books},
  address   = {London},
  year      = {1979}
}
\end{document}